\documentclass[
useAMS,usenatbib,
usegraphicx]{mn2e}

\usepackage{color,soul} 
\usepackage{times} 
\usepackage[
  colorlinks=true, 
  citecolor=blue, 
  linkcolor=red, 
  urlcolor=black,
  dvips]{hyperref}
\usepackage{subfigure}
\usepackage[normalem]{ulem}
\definecolor{green1}{rgb}{0.,0.65,0.}
\makeatletter
\let\jnl@style=\rm
\def\ref@jnl#1{{\jnl@style#1}}

\def\aj{\ref@jnl{AJ}}                   
\def\actaa{\ref@jnl{Acta Astron.}}      
\def\araa{\ref@jnl{ARA\&A}}             
\def\apj{\ref@jnl{ApJ}}                 
\def\apjl{\ref@jnl{ApJ}}                
\def\apjs{\ref@jnl{ApJS}}               
\def\ao{\ref@jnl{Appl.~Opt.}}           
\def\apss{\ref@jnl{Ap\&SS}}             
\def\aap{\ref@jnl{A\&A}}                
\def\aapr{\ref@jnl{A\&A~Rev.}}          
\def\aaps{\ref@jnl{A\&AS}}              
\def\azh{\ref@jnl{AZh}}                 
\def\baas{\ref@jnl{BAAS}}               
\def\bac{\ref@jnl{Bull. astr. Inst. Czechosl.}}
\def\caa{\ref@jnl{Chinese Astron. Astrophys.}}
\def\cjaa{\ref@jnl{Chinese J. Astron. Astrophys.}}
\def\icarus{\ref@jnl{Icarus}}           
\def\jcap{\ref@jnl{J. Cosmology Astropart. Phys.}}
\def\jrasc{\ref@jnl{JRASC}}             
\def\memras{\ref@jnl{MmRAS}}            
\def\mnras{\ref@jnl{MNRAS}}             
\def\na{\ref@jnl{New A}}                
\def\nar{\ref@jnl{New A Rev.}}          
\def\pra{\ref@jnl{Phys.~Rev.~A}}        
\def\prb{\ref@jnl{Phys.~Rev.~B}}        
\def\prc{\ref@jnl{Phys.~Rev.~C}}        
\def\prd{\ref@jnl{Phys.~Rev.~D}}        
\def\pre{\ref@jnl{Phys.~Rev.~E}}        
\def\prl{\ref@jnl{Phys.~Rev.~Lett.}}    
\def\pasa{\ref@jnl{PASA}}               
\def\pasp{\ref@jnl{PASP}}               
\def\pasj{\ref@jnl{PASJ}}               
\def\rmxaa{\ref@jnl{Rev. Mexicana Astron. Astrofis.}}%
\def\qjras{\ref@jnl{QJRAS}}             
\def\skytel{\ref@jnl{S\&T}}             
\def\solphys{\ref@jnl{Sol.~Phys.}}      
\def\sovast{\ref@jnl{Soviet~Ast.}}      
\def\ssr{\ref@jnl{Space~Sci.~Rev.}}     
\def\zap{\ref@jnl{ZAp}}                 
\def\nat{\ref@jnl{Nature}}              
\def\iaucirc{\ref@jnl{IAU~Circ.}}       
\def\aplett{\ref@jnl{Astrophys.~Lett.}} 
\def\apspr{\ref@jnl{Astrophys.~Space~Phys.~Res.}}
\def\bain{\ref@jnl{Bull.~Astron.~Inst.~Netherlands}} 
\def\fcp{\ref@jnl{Fund.~Cosmic~Phys.}}  
\def\gca{\ref@jnl{Geochim.~Cosmochim.~Acta}}   
\def\grl{\ref@jnl{Geophys.~Res.~Lett.}} 
\def\jcp{\ref@jnl{J.~Chem.~Phys.}}      
\def\jgr{\ref@jnl{J.~Geophys.~Res.}}    
\def\jqsrt{\ref@jnl{J.~Quant.~Spec.~Radiat.~Transf.}}
\def\memsai{\ref@jnl{Mem.~Soc.~Astron.~Italiana}}
\def\nphysa{\ref@jnl{Nucl.~Phys.~A}}   
\def\physrep{\ref@jnl{Phys.~Rep.}}   
\def\physscr{\ref@jnl{Phys.~Scr}}   
\def\planss{\ref@jnl{Planet.~Space~Sci.}}   
\def\procspie{\ref@jnl{Proc.~SPIE}}   

\makeatother

\def\lesssim{\mathrel{\hbox{\rlap{\hbox{\lower3pt\hbox{$\sim$}}}\hbox{\raise2pt\hbox{$<$}}}}}
\def\gtrsim{\mathrel{\hbox{\rlap{\hbox{\lower3pt\hbox{$\sim$}}}\hbox{\raise2pt\hbox{$>$}}}}}

\def\phs{\phantom{-}}

\defcitealias{FS90}{1990AJ....100....1F}

\title[Antlia cluster: Scaling Relations]{
  Early-type galaxies in the Antlia Cluster:\\
  A deep look into scaling relations} \author[J. P. Calder\'on
et\,al.]{%
  Juan P. Calder\'on$^{1,2}$
  \thanks{E-mail:\,jpcalderon@fcaglp.unlp.edu.ar}, Lilia
  P. Bassino$^{1,2}$, Sergio A. Cellone$^{1,2}$, \newauthor Tom
  Richtler$^{3}$, Juan P. Caso $^{1,2}$,
  and Mat\'ias G\'omez$^{4}$\\
  $^{1}$Grupo de Investigaci\'on CGGE, Facultad de Ciencias
  Astron\'omicas y Geof\'isicas,
  Universidad Nacional de La Plata, and \\
  Instituto de Astrof\'isica de La Plata (CCT La Plata -- CONICET,
  UNLP),
  Paseo del Bosque S/N, B1900FWA La Plata, Argentina\\
  $^{2}$Consejo Nacional de Investigaciones Cient\'ificas y
  T\'ecnicas,
  Rivadavia 1917, C1033AAJ Ciudad Aut\'onoma de Buenos Aires, Argentina\\
  $^{3}$Departamento de Astronom\'ia, Universidad de Concepci\'on,
  Casilla 160--C, Concepci\'on, Chile \\
  $^{4}$Departamento de Ciencias F\'isicas, Facultad de Ciencias
  Exactas, Universidad Andr\'es Bello, Santiago, Chile}

\begin{document}

\date{...}

\pagerange{\pageref{firstpage}--\pageref{lastpage}} \pubyear{}

\maketitle
\label{firstpage}
\begin{abstract}
  We present the first large-scale study of the photometric and
  structural relations followed by early-type galaxies (ETGs) in the
  Antlia cluster. Antlia is the third nearest populous galaxy cluster
  after Fornax and Virgo (d\,$\sim 35$\,Mpc).  A photographic catalog
  of its galaxy content was built by Ferguson \& Sandage in 1990
  (FS90).  Afterwards, we performed further analysis of the ETG
  population located at the cluster centre. Now, we extend our study
  covering an area four times larger, calculating new {\it total}
  magnitudes and colours, instead of isophotal photometry, as well as
  structural parameters obtained through S\'ersic model fits 
  extrapolated to infinity. Our present work involves a total of 177
  ETGs, out of them 56 per cent have been cataloged by FS90 while the
  rest (77 galaxies) are newly discovered ones. 

  Medium-resolution GEMINI and VLT spectra are used to 
  confirm membership when available. Including radial 
  velocities from the literature, 59\,ETGs are confirmed as Antlia 
  members.  

  Antlia scaling relations mainly support the existence of unique
  functions (linear and curved) that join bright and dwarf ETGs,
  excluding compact ellipticals (cEs). Lenticular galaxies are
  outliers only with respect to the curved relation derived for
  effective surface brightness versus absolute magnitude. The small
  number of bright ellipticals and cEs present in Antlia, prevents us
  from testing if the same data can be fitted with two different
  linear sequences, for bright and dwarf ETGs. However, adding data
  from other clusters and groups, the existence of such sequences is
  also noticeable in the same scaling relations.
\end{abstract}


\section{Introduction}\label{sec:introduction}
Dwarf elliptical (dE) galaxies have been studied extensively from
low-density \citep{2010Ap.....53..462K, 2011MNRAS.416..601S,
  2013ApJ...767..131L} to highly populated environments
\citep{2003AJ....125.1926G, 2008ApJ...679L..77S, 2009MNRAS.393..798D,
  2011MNRAS.410.1076P}. The fact that they are the most abundant
morphological galaxy type in clusters and groups
\citep{1988ARA&A..26..509B, 1998A&A...336...98A}, allows statistically
significant results to be obtained from a thorough analysis of the
early-type population within a given environment.

According to current structure formation models, dwarfs may be
  the descendants of building blocks of larger systems
\citep{1978MNRAS.183..341W}. We are particularly interested in the
formation and evolution discussion \citep{2009ApJ...696L.102J,
    2009ApJS..182..216K, 2012ApJS..198....2K, 2013pss6.book...91G,
    2014MNRAS.443.3381P}, one of whose main points regards whether
there is a link between dwarf (dEs and dSphs) and more luminous
elliptical (E) galaxies.

Different scenarios have been proposed to account for the formation of
early-type galaxies: (i) the monolithic collapse
\citep{1962ApJ...136..748E} in which there was an early major
star-formation burst as a result of the collapse of primordial gas,
producing the most massive galaxies in short periods of time, and the
smallest ones as the universe evolved; (ii) the hierarchical merger
scenario, stating that the minor structures merged to build up the
larger ones. It was proposed by \cite{1977egsp.conf..401T} that
current massive elliptical galaxies are the result from mergers of
disk galaxies.  The mechanism that allows this transformation could be
heating and sweeping out of the galactic gas by supernova-driven winds
and a series of star formation episodes \citep{1987A&A...188...13Y,
  1988MNRAS.233..553D, 2007ApJ...665..265F,
  2009ApJ...690.1452N}. Environmental effects, on the other hand, are
invoked as a means to transform late-type into early-type
galaxies. Among these effects, we can consider starvation
\citep{1980ApJ...237..692L}, galaxy harassment, ram-pressure
stripping, and tidal effects \citep{2001ApJ...559..754M}. In this
sense, the galaxy harassment model proposed by
\cite{1999MNRAS.304..465M} predicts that massive spirals may turn into
lenticular (S0) galaxies due to the loss of their gas, while low-mass
spirals become the current dE through gas loss and kinematic heating
of their stellar disks \citep{2001ApJ...559..791C}.  There is an
important amount of observational evidence that shows similarities
between disk galaxies and dEs, thus supporting this scenario
\citep{2002A&A...391..823B, 2003A&A...400..119D, 2006AJ....132.2432L,
  2011A&A...526A.114T}.

The study of scaling relations followed by galaxies with different
morphologies is a way to explore the evolutionary history of these
systems \citep[and references therein]{2008ASPC..390..403C,
  2013PASA...30...34S}.  While there is overall consensus that both Es
and dEs (but not compact ellipticals, cE) follow the same relation
between luminosity and surface-brightness profile shape \citep[the
  latter quantified by the S\'ersic index $n$;
  see][]{1997ASPC..116..239J}, other scale relations \citep[e.g., the
  Kormendy relation;][]{1985ApJ...295...73K} have originally been
interpreted as evidencing a strict dichotomy between dE and E
galaxies, thus suggesting different origins for them \citep[e.g.:][and
  references therein]{2012ApJS..198....2K}. In opposition, some works
have attempted to show a continuity in scaling relations, which would
imply a continuity of physical properties along the dE -- E sequence
as a signature of a common origin \citep{2003AJ....125.2936G}.

Scaling relations are constructed using either global (effective
radius, $r_\mathrm{e}$; effective surface brightness,
$\mu_\mathrm{e}$) or central (central surface brightness, $\mu_0$)
parameters. \cite{2009ApJS..182..216K} argued that the different
trends in scaling relations between Es and dEs are not due to cores or
extra light in their inner regions. These features contribute a small
percent to total galaxy luminosity and are excluded from the S\'ersic
($r^{1/n}$) fits, which globally match the surface brightness profiles
of early-type galaxies. On the other hand, it would seem that the
  E--dE dichotomy vanishes when $\mu_0$ is measured as the central
  extrapolation of the surface brightness profile, as shown by
  \cite{2003AJ....125.2936G}. This alternative way to understand the
relation between dEs and Es has been developed by
\cite{2011EAS....48..231G} using two linear relations observed in
clusters ($\mu_0$ versus $M$, and $\mu_0$ versus $n$) to derive curved
relations between luminosity and effective parameters, thus turning
the (apparent) dichotomies into continuous relations.

The present work addresses this subject, with the aim of exploring
scaling relations for early-type galaxies in the Antlia Cluster, in
this respect a still mostly unstudied environment.  One main advantage
is the homogeneous CCD photometry of every object in the sample.
Galaxy profile fits using S\'ersic models have been performed in order
to obtain the effective and shape parameters ($r_\mathrm{e}$,
$\mu_\mathrm{e}$, $n$) of the galaxies in the cluster centre and
surrounding areas. In the following, we adopt a distance modulus $m-M
= 32.73$\,mag for Antlia \citep{2003A&A...408..929D}.

We carried out previous CCD studies of the Antlia early--type galaxies
(ETGs) \citep{2008MNRAS.386.2311S, 2012MNRAS.419.2472S}, focused on
those located at the cluster centre.  The photometric techniques used
in those papers were the following: SExtractor automatic measures (for
the majority of low to intermediate luminosity galaxies), and isophote
fits using the \textsc{ellipse} task within
\textsc{IRAF}\footnote{IRAF is distributed by the National Optical
  Astronomy Observatories, which are operated by the Association of
  Universities for Research in Astronomy, Inc., under cooperative
  agreement with the National Science Foundation.} (for the brightest
objects). In both cases, magnitudes and colours were measured up to a
fixed isophotal radius.

In the present paper, we extend the study of the Antlia ETGs including
adjacent regions, covering an area four times larger and performing a
new {\it total} photometry, extrapolating S\'ersic models to infinity.
We want to remark that Antlia has a particular structure, with two
dominant galaxies (NGC\,3258 and NGC\,3268), which has been
interpreted by means of X--ray data as a galaxy cluster in an
intermediate merger stage \citep{2011HEAD...12.3903H,
  1997ApJ...485L..17P, 2000PASJ...52..623N}. It thus provides us with
the opportunity to test scaling relations in an environment where
significant pre-processing should have taken place.  We also aim at
positioning the Antlia cluster scaling relations in the current
picture along with already studied groups and clusters.

This work is organized as follows. In Section
\ref{sec:observations-and-surface-photometry} we describe the data and
summarize the reduction processes. Section \ref{sec:sample-selection}
addresses the galaxy selection.  A discussion is given in Section
\ref{sec:results-and-discussion}, including the scaling relations
obtained with photometric and structural parameters as well as a
comparison with other systems. Finally, the conclusions are given in
Section \ref{sec:summary-and-conclusions}.

\begin{table*}
\begin{minipage}{110mm}
  \caption{Observational data for the MOSAIC\,II fields. The column
    $n_\mathrm{f}$ is the number of exposures that have been
    averaged on each field shown in Fig\,\ref{fig:fields}.}
\label{tab:fields-coordinates}
\begin{tabular}{@{}ccccccccc}
\hline
Field & Observation & $\alpha_{2000}$ & $\delta_{2000}$ &
                       Filter & $n_\mathrm{f}$ & Exposure & Airmass & FWHM \\
      & date        &                    &                   &
      &             &  [seconds]         &                 & [\arcsec] \\
\hline
0 & April 2002 & 10:29:22 & $-$35:27:54 & $R$ & 5 &  600 & 1.059 & 1.0 \\
  &            &          &             & $C$ & 7 &  600 & 1.037 & 1.1 \\
1 & March 2004 & 10:28:59 & $-$34:57:40 & $R$ & 5 &  600 & 1.588 & 1.0 \\
  &            &          &             & $C$ & 7 & 1000 & 1.076 & 1.0 \\
2 & March 2004 & 10:31:09 & $-$34:55:59 & $R$ & 5 &  600 & 1.056 & 1.0 \\
  &            &          &             & $C$ & 7 &  900 & 1.016 & 1.2 \\
3 & March 2004 & 10:31:35 & $-$35:30:42 & $R$ & 5 &  600 & 1.269 & 0.9 \\
  &            &          &             & $C$ & 7 &  900 & 1.030 & 0.8 \\
\hline
\end{tabular}
\end{minipage}
\end{table*}

\section{Observations and Surface
  photometry}\label{sec:observations-and-surface-photometry}
\subsection{Data}\label{sec:data}
\begin{figure*}
 \includegraphics[scale=0.35]{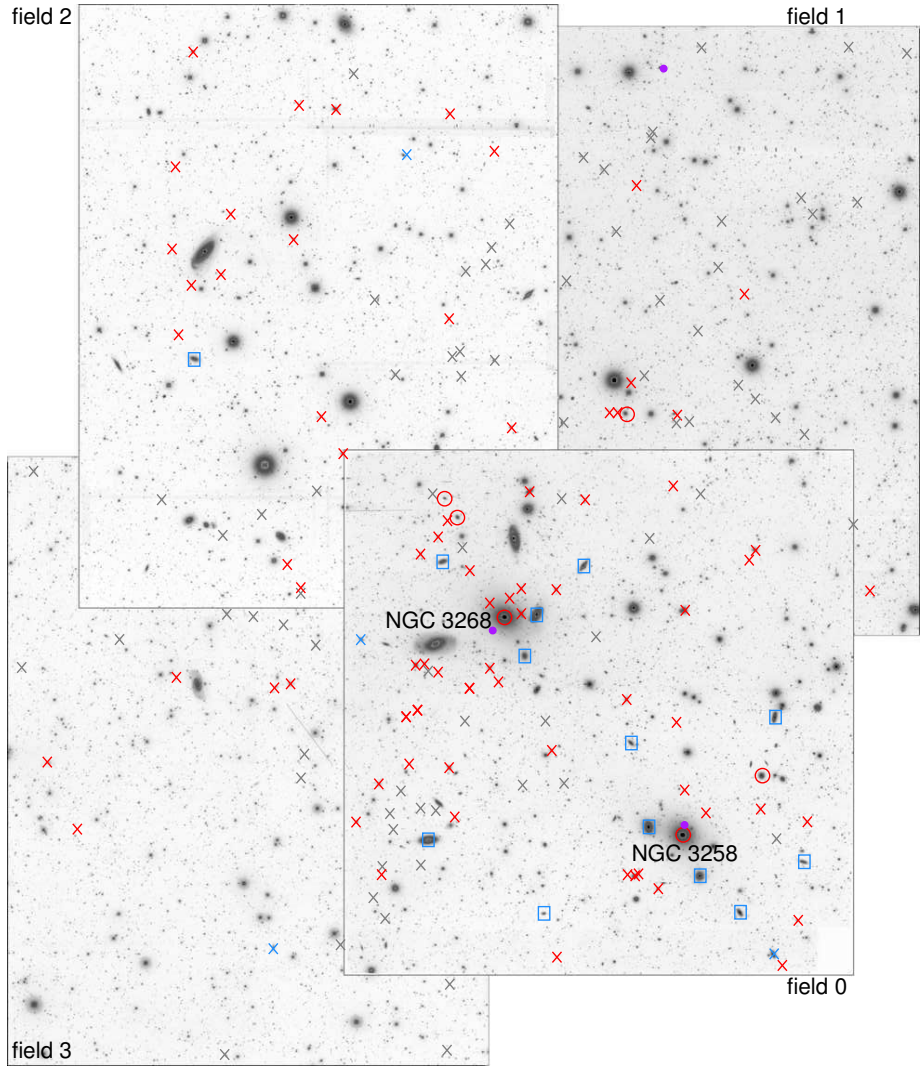}
 \caption{Composite $T_1$ image of the Antlia cluster (four
   MOSAIC\,II fields). North is up and east to the left. Red open
   circles identify E galaxies, red crosses dEs, blue open squares
   S0s, blue crosses dS0s, purple filled circles cEs, and grey crosses
   new uncatalogued galaxies. We use the same symbols in the rest of
   the paper.}
  \label{fig:fields}
\end{figure*}

We perform a photometric study with images of the Antlia cluster from
two observing runs, obtained with the MOSAIC\,II wide-field camera
mounted at the 4-m Blanco telescope of the Cerro Tololo Inter-American
Observatory (CTIO, Chile). During the first run (April 2002) we
observed the cluster centre (hereafter field 0), while in the second
run (March 2004) we observed three adjoining fields located at the N,
NE and E of the central one (hereafter fields 1, 2 and 3,
respectively). Each MOSAIC\,II field covers 36\,arcmin $\times$
36\,arcmin, that corresponds to about $370 \times 370$\,kpc$^2$
according to the adopted Antlia distance. Field 0 includes the two
giant elliptical galaxies (gEs) located at the cluster centre
(NGC\,3258 and NGC\,3268). Within our Antlia Cluster Project, this
central field has been used to study the globular cluster systems of
these two galaxies \citep{2003A&A...408..929D}, to perform the first
CCD analysis of the ETG population at the cluster centre
\citep{2008MNRAS.386.2311S, 2012MNRAS.419.2472S}, as well as to
investigate the ultracompact dwarfs
\citep[UCDs,][]{2013MNRAS.430.1088C, 2014MNRAS.442..891C}. In addition
to this central field, the three adjacent fields are used for the
present study. Fig.\,\ref{fig:fields} shows a composition of the four
fields.  They overlap with each other in order to obtain a homogeneus
photometry.

We used the Kron-Cousins $R$ and Washington $C$ filters, and the
instrumental magnitudes were later transformed into those
corresponding to the genuine Washington $C$ and $T_1$ bands. The
  MOSAIC\,II images were reduced using the \textsc{mscred} package
  within \textsc{IRAF}. Each image was processed using
  \textsc{ccdproc}, applying the overscan, bias level and flat field
  corrections. The individual MOSAIC extensions were then combined
  into a single FITS image using the \textsc{msccmatch},
  \textsc{mscimage} and \textsc{mscimatch} tasks, which were used to
  scale to a common flux level, match coordinates and adjust WCS
  (World Coordinate System), and also the cosmic ray correction was
  performed. We next subtracted a second order polynomial
  surface from the background using \textsc{mscskysub} (when
  necessary) to remove any residual large-scale gradients. In each
  filter, the images were aligned and a single stacked image was
  created using \textsc{mscstack} with \textsc{ccdclip} pixel
  rejection. More details on the images' reduction will be given in a
  forthcoming data paper (Calder\'on et al., in preparation).

Table \ref{tab:fields-coordinates} gives the observing log, including
the dates, position of the field centres, filter, number of exposures
($n_\mathrm{f}$) that have been averaged to obtain the final image for
each field, exposures, mean airmass, and seeing for these final
images.

Regarding the calibration to the standard system, we used the
transformation equations for the central field (0) given by
\cite{2003A&A...408..929D}. For the three adjoining fields (1, 2 and
3), we obtained the following relations between instrumental and
standard magnitudes based on standard stars fields from the list of
\cite{1996AJ....111..480G},

\begin{eqnarray}\label{eq:calibration}
T_1&=&(m_R~+~0.02)~+~a_1~+~a_2~X_{R}~+~a_3~(C-T_1)\\
C&=&m_C~+~b_1~+~b_2~X_C~+~b_3~(C-T1) \nonumber\textnormal{,}
\end{eqnarray}
where the coefficients and their errors are given in Table
\ref{tab:calibration-errors}, $m_R$, $m_C$ are the instrumental
magnitudes and $X_R$, $X_C$ the respective airmasses.  All the
magnitudes and colours presented in the figures of this paper have
been corrected for Galactic absorption and reddening. The colour
excess $E(B-V)$ was provided by \cite{2011ApJ...737..103S}, and we
used the relations $E(C-T_1)=1.97\, E(B-V)$
\citep{1977AJ.....82..798H} and $A_R/A_V=0.75$
\citep{1985ApJ...288..618R} to obtain the absorption and reddening in
the Washington system.

As a consequence of using images taken in two different runs (two
years apart) in combination with the large size of the MOSAIC\,II
field, differences in the zero point magnitudes, for each filter and
between the four fields, are expected.  In order to estimate such
offsets, we calculated $C$ and $T_{1}$ magnitudes of the point-sources
that lie in the overlapping areas and computed the respective mean
differences between fields 1, 2, and 3 with respect to the central (0)
one. Finally, we applied the zero point offsets, and referred all $C$
and $T_{1}$ magnitudes to the central field. The offsets are
  higher in the $T_{1}$ band ($\approx 0.1$\,mag), while in $C$-band
  are between $0.01$ y $0.04$\,mag.

In addition to the MOSAIC\,II images, we have medium-resolution
spectra obtained with GEMINI-GMOS (programmes GS-2011A-Q-35 and
GS-2013A-Q-37) and VIMOS-VLT (programme 79.B-0480), all of them in
multi-object mode. A description of the observations, reduction, and
radial velocity measurement corresponding to the GMOS spectra are
given by Caso et al. (2014; a paper dealing with UCDs and bright GCs
of NGC\,3268), while those to the VIMOS spectra are given by Caso et
al. (in preparation; a global kinematic study of the Antlia cluster).

\subsection{Surface photometry}\label{sec:surface-photometry}
We used the \textsc{ellipse} task within the \textsc{isophote} IRAF's
package, to obtain the observed surface brightness profiles (surface
brightness versus equivalent radius, being $r = \sqrt{a~b} = a
\sqrt{1-\epsilon}$, where $a$ is the isophote semi-major axis and
$\epsilon$ its ellipticity) of all ETGs in the sample. In every case,
we set off to test the elliptical fits without fixing the geometric
parameters, such as ellipticity, position angle, centre coordinates,
etc.  In some cases, a completely free parameter model could not be
fitted due to different reasons like images defects, extremely weak
objects, nearby saturated stars. Then, we improved the fits as much as
possible, fixing some of the parameters and/or changing the fitting
step. For every galaxy, we also built bad-pixels masks to
  flag-out pixels from the fits and avoid contaminating background or
  foreground objetcs.

\begin{table}
\begin{minipage}{\columnwidth}
\begin{center}
  \caption{Coefficients and errors of the calibration equations
   in Eq. \ref{eq:calibration}.}
\label{tab:calibration-errors}
\begin{tabular}{@{}ccccccc}
  \hline
  & $a_1$ & $a_2$ & $a_3$ & $b_1$ & $b_2$ & $b_3$ \\
  \hline
  Coeff. & 0.608 & $-0.140$   & 0.0184 & $-0.059$   & $-0.418$   & 0.111 \\
  Error & 0.003  & $\phs 0.001$ & 0.0020 & $\phs 0.004$ & $\phs 0.006$ & 0.005 \\
  \hline
\end{tabular}
\end{center}
\end{minipage}
\end{table}

Due to the large MOSAIC\,II field, we preferred to estimate the
background (sky level) for each galaxy independently, instead of
setting the same background level for the whole image. We first
calculated an initial value taking the mode from several positions,
free of other sources, around the galaxy. Afterwards, we applied an
iterative process until the outer part of the integrated flux profile,
i.e. for large galactocentric distances, became as flat as possible.
The details of the background estimation will be given in the future
data paper.

Regarding the SN ratio, we followed \cite{2011MNRAS.414.2055M} and
calculated the signal-to-noise ratio at the
$27.5$\,magnitude\,arcsec$^{-2}$ isophote for both bands. For the
faintest objects ($T_{1} > 14$\,mag) the S/N spans a range from $1.6
\pm 0.3$ to $3.0 \pm 1.0$ in the $R$ filter; while in the $C$ filter
the range is from $3.2 \pm 2.0$ to $5.6 \pm 2.1$. Note that, despite
the lower S/N, the $27.5$\,magnitude\,arcsec$^{-2}$ isophote
corresponds to a larger physical radius in $T_{1}$ than in $C$, so the
galaxies are more clearly detected in $T_{1}$.

In order to test the \textsc{ellipse} output, we varied the geometric
parameters over a small range to confirm if the final observed profile
converged to a unique solution. In addition, we used different steps
for each galaxy to check how to obtain the best observed profile.

Then, we fitted the S\'ersic model \citep{1968adga.book.....S} to
every observed galaxy surface brightness $C$ and $T_1$ profiles using,
\begin{equation}\label{eq:sersic_law}
  \mu(r) = \mu_\mathrm{e} + 1.0857 \cdot b_n \left[ \left(\frac{r}{r_\mathrm{e}}\right)^{1/n} - 1 \right]\textnormal{,}
\end{equation} 
where $r_\mathrm{e}$ is the effective radius, $\mu_\mathrm{e}$ is the
effective surface brightness, and $n$ is the S\'ersic shape index that
is a measure of the concentration of the light profile. The function
$b_n$ depends on the shape parameter $n$ and we applied a numerical
method to obtain it by solving the equation
\citep{1991A&A...249...99C},
\begin{equation}\label{eq:bene}
\frac{\Gamma(2n)}{2} = \gamma(2n,b_n)\textnormal{,}
\end{equation} 
where $\Gamma(x)$ is the complete gamma function and $\gamma(a,x)$ the
incomplete gamma function.

The S\'ersic model is integrated to infinity to obtain {\it total}
magnitudes and colours. The integrated magnitude results,
\begin{eqnarray}\label{eq:sersic_int}
m &=&  \mu_\mathrm{e} - 1.99 - 5 \log(r_\mathrm{e}) - 1.0857 \, b_n-  \\
& & 2.5 \log\left[b_n^{-2n}\,n\,\Gamma(2n)\right] \nonumber\textnormal{.}
\end{eqnarray}
The profile fits were carried out with the \textsc{nfit1d} task within
IRAF, that performs a $\chi^2$ residual minimisation using the
Levenberg-Marquardt algorithm. The inner 1\,arcsec of the profiles was
not included in the fits to avoid seeing effects. Anyway, we will show
in the data paper that the seeing does not affect the profile fits for
galaxies with index $n \lesssim 4$.

In most cases, we were able to fit the ETG profiles with only one
S\'ersic model with residuals below 0.2\,mag\,arcsec$^{-2}$.  We
want to make clear that the parameters used for the scaling relations
were derived without attempting any bulge-disk decomposition. This may
seem particularly inappropriate for S0 galaxies, but note that also
cEs and gEs, along with a significant fraction of (bright) dEs, do
show two-component profiles even when a clear bulge-disk distinction
cannot be made. Our approach thus traces the overall morphology of the
ETGs, including both components, when two are present. For
  example, a decomposition analysis of Virgo early-type dwarf galaxies
  has been performed by \cite{2011MNRAS.414.2055M} and
  \cite{2014ApJ...786..105J}.

\section{Sample selection}\label{sec:sample-selection}
In this section we describe how the galaxy sample was formed. The
original source was the \citet[hereafter FS90]{1990AJ....100....1F}
Antlia catalogue. We used their coordinates to identify all the
catalogued galaxies on our MOSAIC\,II fields.  Also, we added the
galaxies discovered in our previous study of the central field
\citep{2012MNRAS.419.2472S}, and now we include new candidates located
in the four fields that have not been catalogued before.

\subsection{Original sample}\label{sec:original-sample}
The first large-scale study of the galaxy population of the Antlia
cluster is the FS90 photographic catalogue, with a limiting magnitude
$B_\mathrm{T} = 18$ ($M_B = -14.7$). On the basis of morphological
criteria they give a membership status (1: `definite' member, 2:
`likely' member, and 3: `possible' member) for each galaxy. There are
375 galaxies (of all types) located over the whole cluster area in
this catalogue. Out of them, 36.5 per cent have membership status 1,
i.e. the highest membership probability, 26.5 per cent status 2, and
the rest (37 per cent) status 3. Only 6 per cent had measured radial
velocities at that time. In successive papers, we obtained IMACS,
GMOS, and VIMOS spectra that, including the velocities published in
NED\footnote{This research has made use of the NASA/IPAC Extragalactic
  Database (NED) which is operated by the Jet Propulsion Laboratory,
  California Institute of Technology, under contract with the National
  Aeronautics and Space Administration.}, let us confirm now $\sim
30$\,per cent of the FS90 ETG galaxies as cluster members. We recall
that to be a confirmed Antlia member, the radial velocity should lie
in the range $1200 - 4200$ km\,s$^{-1}$ \citep{2008MNRAS.386.2311S}.

In particular, considering the galaxies with membership status 1
assigned by FS90 (`definite' members) that have measured radial
velocities from VIMOS spectra, we can confirm that $\approx 96$ per
cent of them are in fact members of the cluster. The high reliability
of the FS90 membership classification has already been pointed out in
previous works \citep[e.g.][and references
therein]{2012MNRAS.419.2472S}. Thus, we will consider the FS90 ETGs
with membership 1 as true members. In that way, all ETGs
confirmed with radial velocities plus those from FS90 with membership
1 will be considered in the rest of this paper as `Antlia members'.

\begin{figure}
     \begin{center}
        \subfigure[FS90 273: Spiral structure.]{%
            \label{fig:FS90273e}
            \includegraphics[width=0.4\columnwidth]{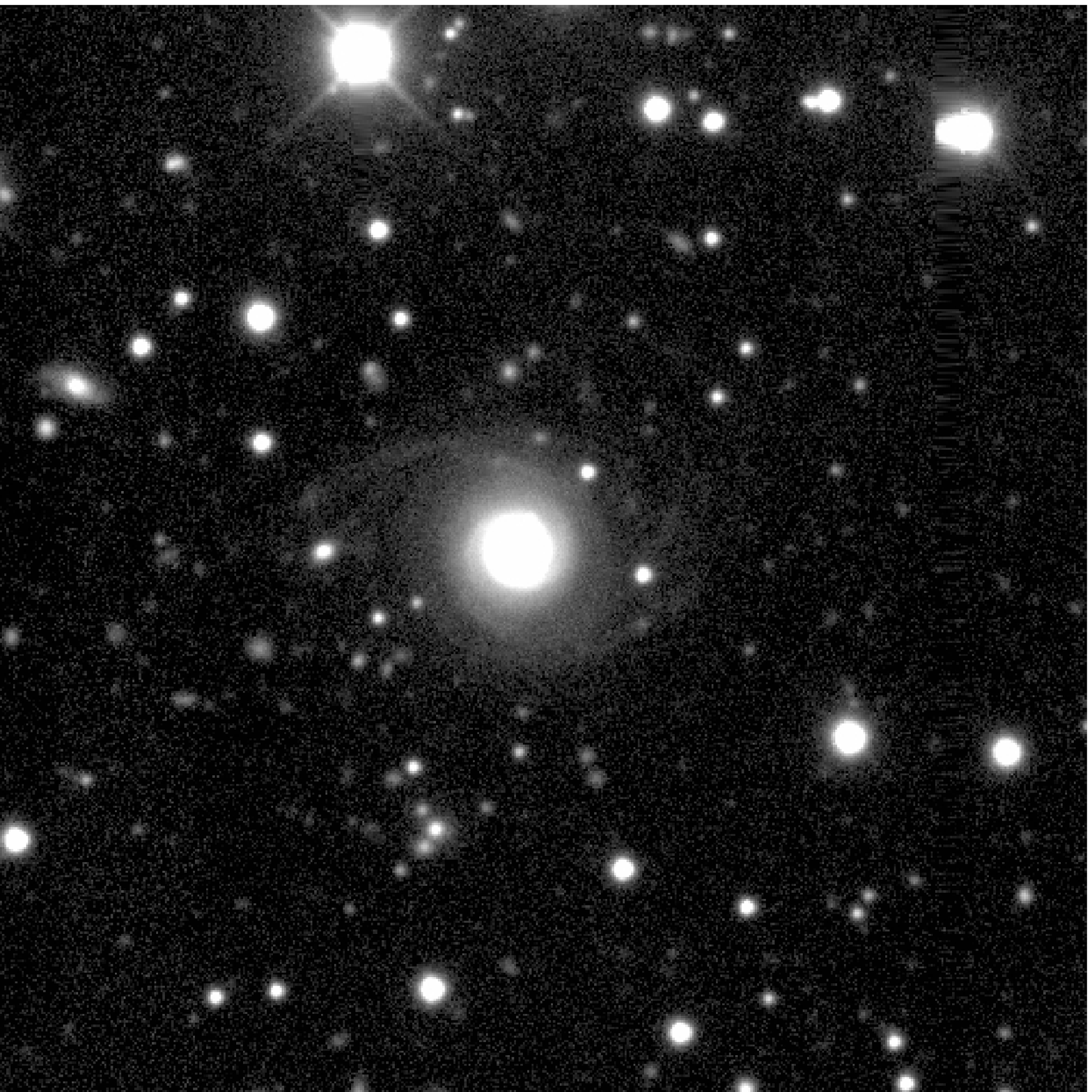}
        }\hspace{0.1\columnwidth}%
        \subfigure[FS90 297: Irregular morphology.]{%
           \label{fig:FS90297e}
           \includegraphics[width=0.4\columnwidth]{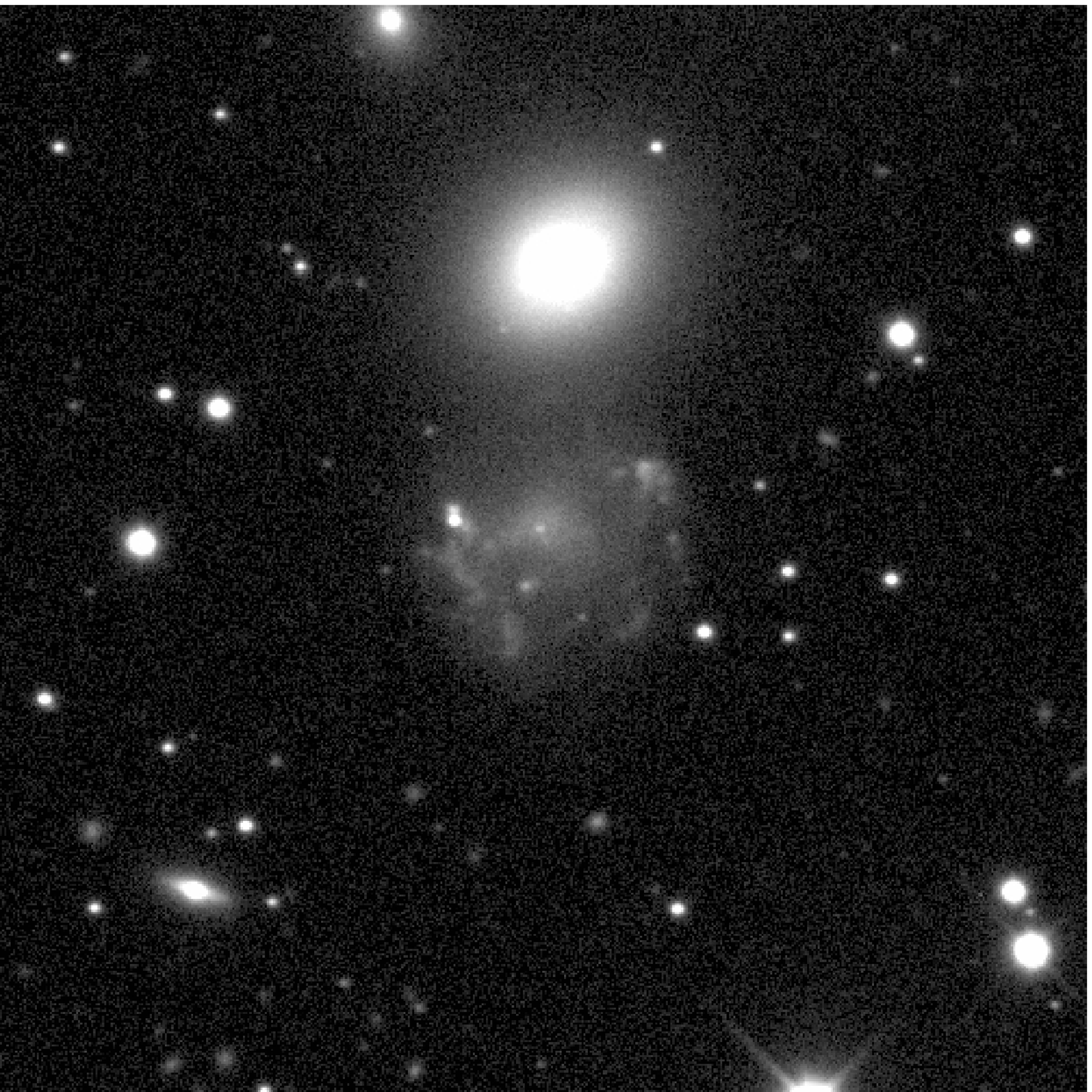}
        }\\
        \subfigure[FS90 283: Nearby bright star.]{%
            \label{fig:FS90283e}
            \includegraphics[width=0.4\columnwidth]{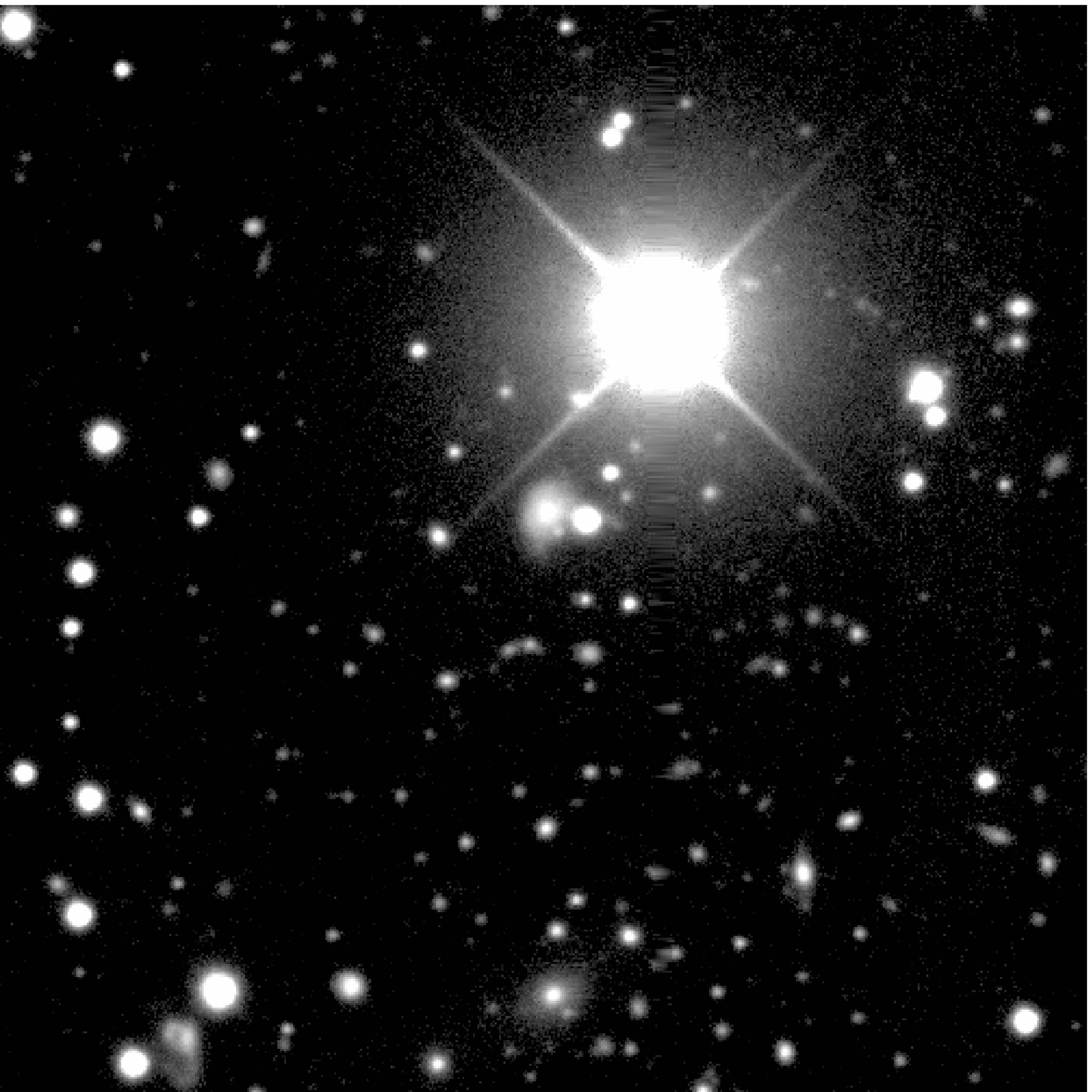}
          }\hspace{0.1\columnwidth}%
          \subfigure[FS90 288: Faint galaxy, overlapping with a bright
          star's spike.]{%
           \label{fig:FS90288e}
           \includegraphics[width=0.4\columnwidth]{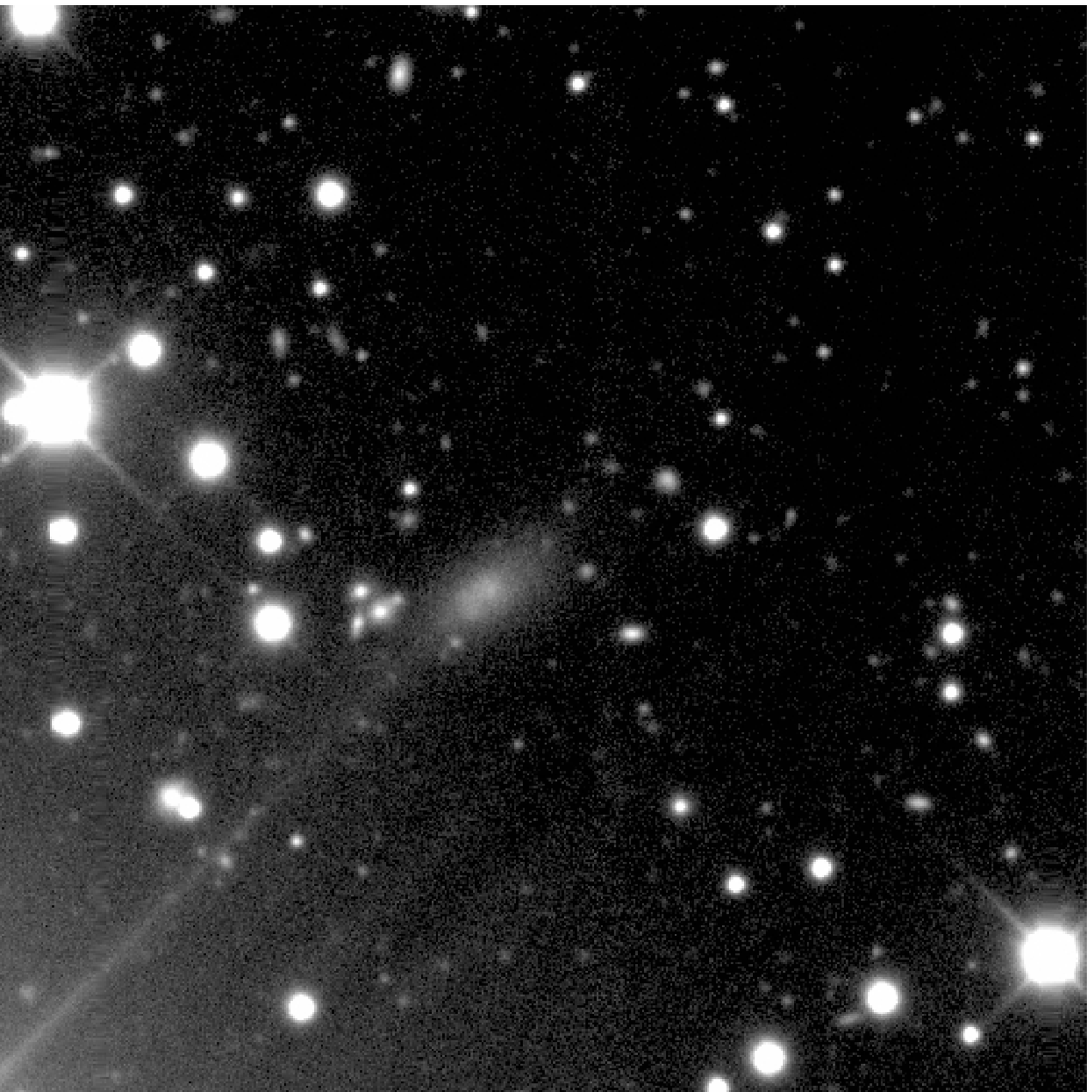}
        }
    \end{center}
    \caption{%
      Examples of galaxies excluded from the sample.  Each image
      covers a field of $23 \times 23$ kpc$^2$ at the Antlia
      distance, being the galaxy located at the centre.  
    }%
   \label{fig:exclude}
\end{figure}

According to this selection, we have obtained surface brightness
profiles in $C$ and $T_1$ for 100 ETGs. Out of them, 53 are
spectroscopically confirmed members. Among the FS90 galaxies, we were
unable to obtain several profiles because the galaxies are very faint
or superimposed on image defects or gaps.  All objects with evident
background morphology and/or central bars were also excluded from the
present study as well as irregular or other star-forming galaxies,
e.g. blue compact dwarfs
\citep[BCDs,][]{2014A&A...563A.118V}. Fig.\,\ref{fig:exclude} shows a
few examples of galaxies excluded from the present sample.

\subsection{Final sample}\label{sec:final-sample}
After a careful visual inspection of $C$ and $T_1$ images of the
four MOSAIC\,II fields, we discovered 77 new ETG candidates that have
not been catalogued before. The preliminary selection was performed
according to the following criteria:

\begin{enumerate}
\item[a.] The galaxy is not affected by bleeding or diffraction due to a
  saturated nearby star, or any cosmetic image defects.

\item[b.] It preserves isophotal shape and compatible \textsc{ellipse}
  outputs in both filters.

\item[c.] The \textsc{ellipse} output attains a reasonable S/N
    ratio out to the $\sim 27.5$ mag\,arcsec$^{-2}$ ($R$--band)
    isophote. There are no objects to mask near the centre of the
    target galaxy.

\item[d.] The residuals of the best S\'ersic fit(s) show no sign of
  spiral structure.

\item[e.] The fit is stable even when performing small changes in the
  initial geometric parameters or the sky level.

\item[f.] We only include in the sample objects with $r \geq
  10$\,arcsec, thus preventing spurious detections.

\item[g.] The new objects have to be identified in both bands, R and
  C, to be added to the sample (and our catalogue).
\end{enumerate}

The fraction of all galaxies detected only on the R images is less
than 5 percent, while no galaxies were detected only on the C
images. This is a consequence of the fact that, for a given surface
brightness level, the S/N is better in the C band, but the size of the
corresponding isophote is larger in the R. The colour bias against the
faint red galaxies should then not be highly relevant.

The newly identified ETG candidates are in the faint magnitude regime
and correspond mainly to morphologies in accordance with dE (nucleated
and non-nucleated) as well as dSph galaxies. For these faint galaxies
the probability of lying in the background is obviously higher than
for the brighter ones, so only part of them may be real cluster
members. As we lack spectra for them, we neither can identify
those that ---even being Antlia members--- should be classified
as late-type instead of ETGs.  As a consequence, adding all these new
galaxy candidates would lead to blur the scaling relations that are
the goal of this paper. Thus, we decided to keep those that have
higher chances of being ETG members, on the basis of the
colour--magnitude relation (CMR) followed by the ETG `members'.

It is known that the CMR of ETGs in galaxy clusters is a well defined
correlation \citep[e.g.][]{2008MNRAS.383..247P, 2011MNRAS.410..280J,
  2012AAS...21941106M}.  Moreover, this CMR (also called `red
sequence') is a universal relation with very small scatter, that
carries information about the formation of the clusters
themselves. The slope of the CMR has been understood as a
mass--metallicity relation, while the effect of the (slightly)
different ages at each stellar mass is most likely causing the small
scatter about the CMR. Such small scatter can be used as an additional
constraint, besides morphology, to identify probable cluster members
in the cases where no spectroscopy is available.  In previous papers
we have shown that this scatter does not increase towards fainter
magnitudes, remaining almost constant along the whole CMR
\citep{2008MNRAS.386.2311S,2012MNRAS.419.2472S}.

\begin{figure}
 \includegraphics[width=\columnwidth]{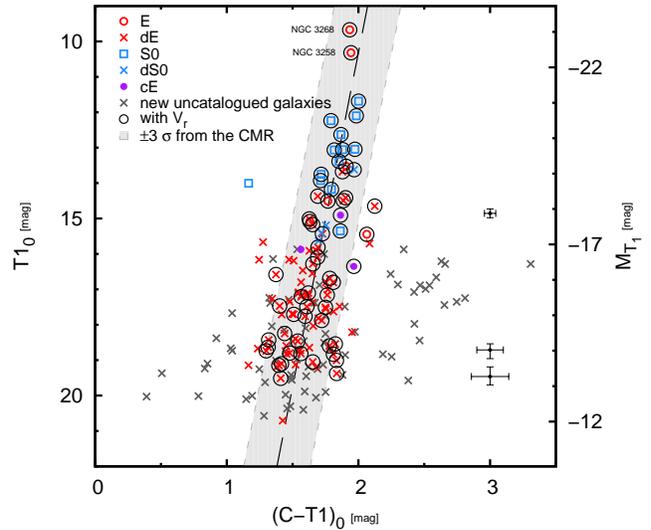}
 \caption{Colour--magnitude diagram of early-type plus newly
   discovered galaxies in the Antlia cluster (symbols are identified
   in the plot). The dashed line shows the CMR followed by the ETG
   `members' (i.e. FS90 ETGs spectroscopically confirmed or with
   membership status 1) while the shaded band identifies the locus
   within $\pm 3 \sigma$ of such relation. The two gEs are identified
   at the bright magnitude end. On the right we show typical errors
   for different magnitudes.}\label{fig:CT1_T1}
\end{figure}

Fig.\,\ref{fig:CT1_T1} shows the CMR followed by the Antlia ETG
`members'. The new uncatalogued galaxies, plotted as grey crosses,
have no radial velocity and cover the fainter half of the diagram,
towards bluer and redder colours with respect to the CMR. The
  real nature of these objects is doubtful (the reddest ones are most
  probably background objects). We decided to add to the final ETG
sample only the new galaxies located within $\pm 3 \sigma$ of the CMR,
that is indicated in the plot as the shaded band. The dispersion
$\sigma$ is calculated with respect to both variables, colour and
magnitude. We stick to the $\pm 3 \sigma$ limit because it is
supported by the fact that member galaxies confirmed with radial
velocity fall within such limits in the CMD (or are located very
close).

Seven of these new galaxies
%
%
%
have VIMOS spectra (Caso et al., in preparation) and their
radial velocities are in the range corresponding to the confirmed
members, as explained in Sec. 3.1. Thus, they will be included
in the `members' sample. Their basic data are presented in
Table\,\ref{tab:new-dwarf-members}, i.e. coordinates and Washington
photometry. They are named like previously discovered new members
\citep{2012MNRAS.419.2472S}, with the acronym ANTL followed by the
J2000
coordinates\footnote{\url{http://cdsweb.u-strasbg.fr/vizier/Dic/iau-spec.htx}}.

\begin{table*}
\begin{minipage}{110mm}
\begin{center}
  \caption{New Antlia dwarf galaxy members confirmed with VIMOS
    spectra.}
\label{tab:new-dwarf-members}
\begin{tabular}{@{}ccccc}
\hline
 ID & $\alpha_{2000}$ & $\delta_{2000}$ & $(T_1)_0$& $(C-T_1)_0$  \\
\hline
ANTL\,J103046.7$-$353918.0 & 10:30:46.7 & $-35$:39:18.0 & 18.24 & 1.43 \\
ANTL\,J103036.4$-$353047.2 & 10:30:36.4 & $-35$:30:47.2 & 17.87 & 1.72 \\
ANTL\,J103033.1$-$352638.4 & 10:30:33.1 & $-35$:26:38.4 & 18.73 & 1.29 \\
ANTL\,J103037.4$-$352708.3 & 10:30:37.4 & $-35$:27:08.3 & 19.10 & 1.41 \\
ANTL\,J103021.4$-$353105.2 & 10:30:21.4 & $-35$:31:05.2 & 15.01 & 1.62 \\
ANTL\,J103022.0$-$353805.3 & 10:30:22.0 & $-35$:38:05.3 & 17.46 & 1.40 \\
ANTL\,J103013.7$-$352458.6 & 10:30:13.7 & $-35$:24:58.6 & 19.37 & 1.83 \\
\hline
\end{tabular}
\end{center}
\end{minipage}
\end{table*}

From now on, we will identify as `final sample' the group comprised of the
107 `member' ETGs plus 31 galaxies from the new ETG sample that lie within
$\pm 3 \sigma$ of the CMR. That is, the final sample contains 138 galaxies.


\section{Results}\label{sec:results-and-discussion}
We will address in the following the scaling relations obtained with the 
final sample of ETGs in the Antlia cluster. In Section
\ref{sec:final-sample} we have already introduced the CMR of ETGs and
how it has been used to select the most probable early-type members
among the newly identified galaxies. We will come back first to the
CMR for a more detailed analysis, and will continue with the scaling
relations that involve global structural parameters of the final
sample.

\subsection{The Colour--Magnitude
  Relation}\label{sec:color-magnitude-relation} As said above,
Fig.\,\ref{fig:CT1_T1} shows the colour--magnitude diagram of the ETGs
in the final sample, with the corresponding morphological types
indicated in the plot. The dashed black line shows the CMR calculated
through a least squares fit of the ETG `members', taking into account
errors in both axes, that gives:
\begin{equation}\label{eq:regression}
  (T_1)_0 = (-18.9 \pm 0.1)~(C-T_1)_0 + (48.1 \pm 2.6) \textnormal{,}
\end{equation}
where the standard deviation $\sigma$ is 1.59. Almost all galaxies that
are spectroscopically confirmed members, identified in
Fig.\,\ref{fig:CT1_T1} with black open circles, fall within $\pm 3
\sigma$ from such CMR. This gives support to our selection of new
galaxies within such colour limits.

In particular, the existence of a 'break' (in the
  sense of slope change) of the bright end of the CMR with respect to
the linear fit performed on all the ETGs was not so evident in our
previous work on the Antlia cluster \citep{2008MNRAS.386.2311S,
  2012MNRAS.419.2472S}. Small differences in the $(C - T_1)$
  colours of the brightest galaxies are probably responsible of this
  `break' at $M_{T_1} \approx -20$\,mag.  In fact, the colours of
  these brightest galaxies are bluer in the present paper ($\Delta
  (C-T1)=0.1$\,mag, 5 galaxies), where magnitudes are not just
  isophotal but {\it total}. It seems that, as a consequence of
  integrating the fitted S\'ersic law to infinity, a slight break can
  be perceived for the first time, at the bright end of the CMR
  (Fig.\,\ref{fig:CT1_T1}). As the Antlia cluster has mainly bright
  lenticulars and few bright ellipticals, this effect is shown by just
  a small number of galaxies.
\begin{figure}
 \includegraphics[width=\columnwidth]{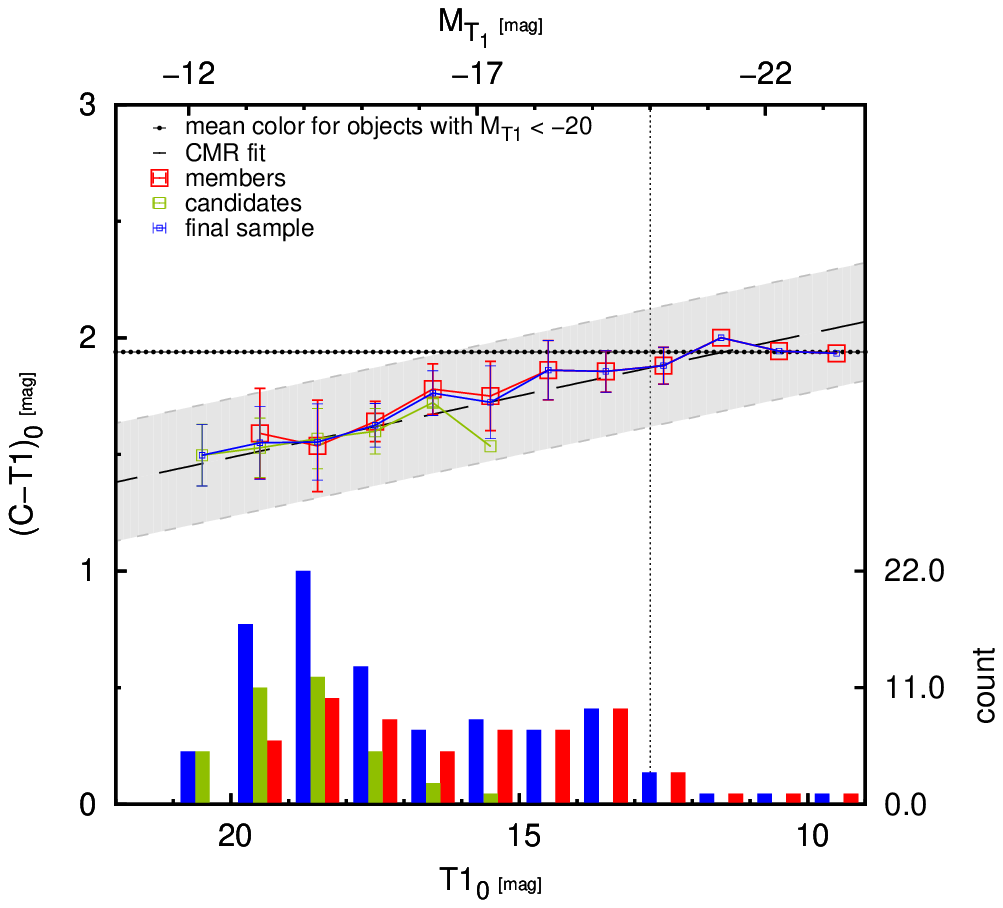}
  \caption{Break at the bright end of the Antlia CMR. 
Top: colour versus magnitude relation, shaded band as 
in Fig.\,\ref{fig:CT1_T1}. Bottom: histogram of galaxy counts 
as a function of magnitude (ordinates on the right-hand side)}
  \label{fig:T10_CT10promedio}
\end{figure}

Fig.\,\ref{fig:T10_CT10promedio} shows an alternative display of the
CMR followed by all the galaxies of the `final sample', i.e. the
`members' and the new galaxies that lie within $\pm 3 \sigma$ of the
CMR. The dashed line is the same fit to the CMR for confirmed members,
over the full range of magnitudes. The horizontal dotted line
represents the mean colour for the brighter end of the relation (i.e.
with $M_{T_1} \le -20$\,mag). The connected points represent the mean
colour of the magnitude bins. It can be seen that for the brighter
end, the colour remains fixed at $(C-T_1)_0 \sim 1.9$. At the bottom,
we present a histogram of the number of galaxies in each magnitude
bin.

\begin{figure*}
  \begin{center}
      \subfigure[Central surface brightness versus absolute magnitude 
      ($V$-band).]{%
       \label{fig:Mv_mu0}
       \includegraphics[width=0.31\textwidth]{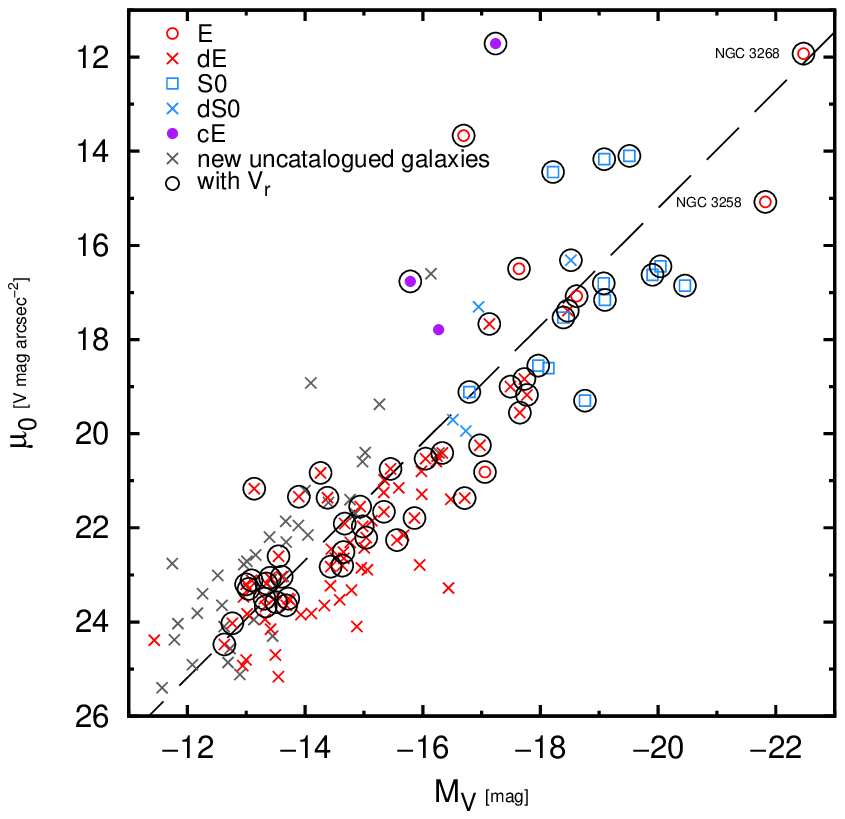}}\hspace{0.015\textwidth}
     \subfigure[Absolute magnitude versus logarithm of the S\'ersic
     index ($V$-band).]{%
       \label{fig:Mv_n}
       \includegraphics[width=0.31\textwidth]{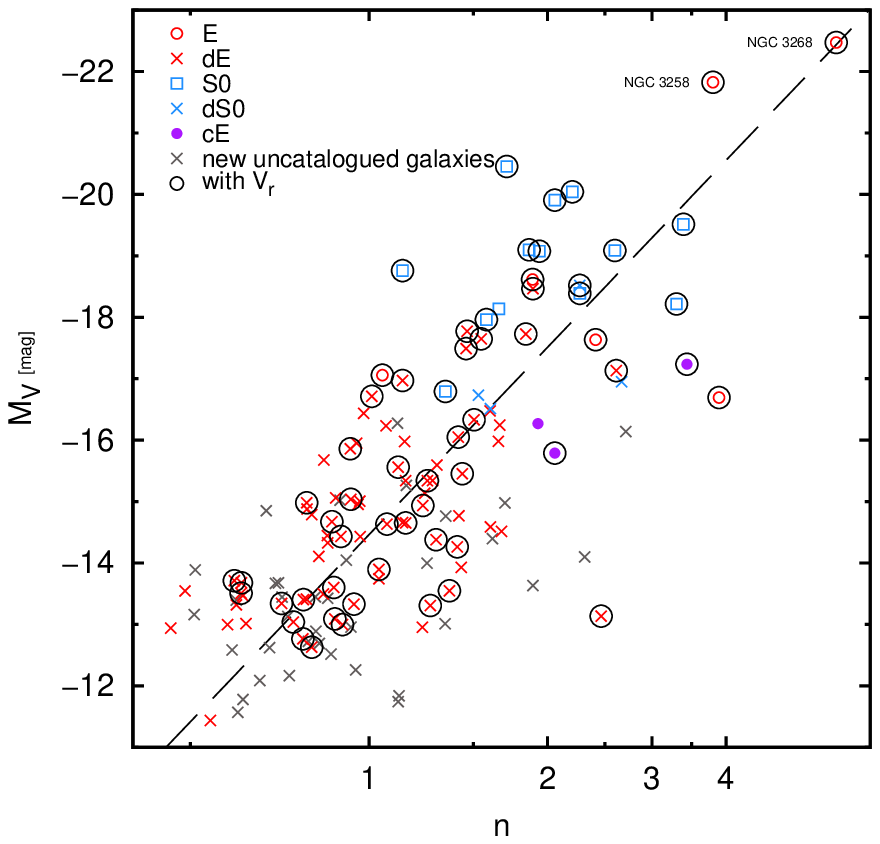}}\hspace{0.015\textwidth}
     \subfigure[Central surface brightness versus logarithm of the
     S\'ersic index ($V$-band).]{%
       \label{fig:mu0_n}
       \includegraphics[width=0.31\textwidth]{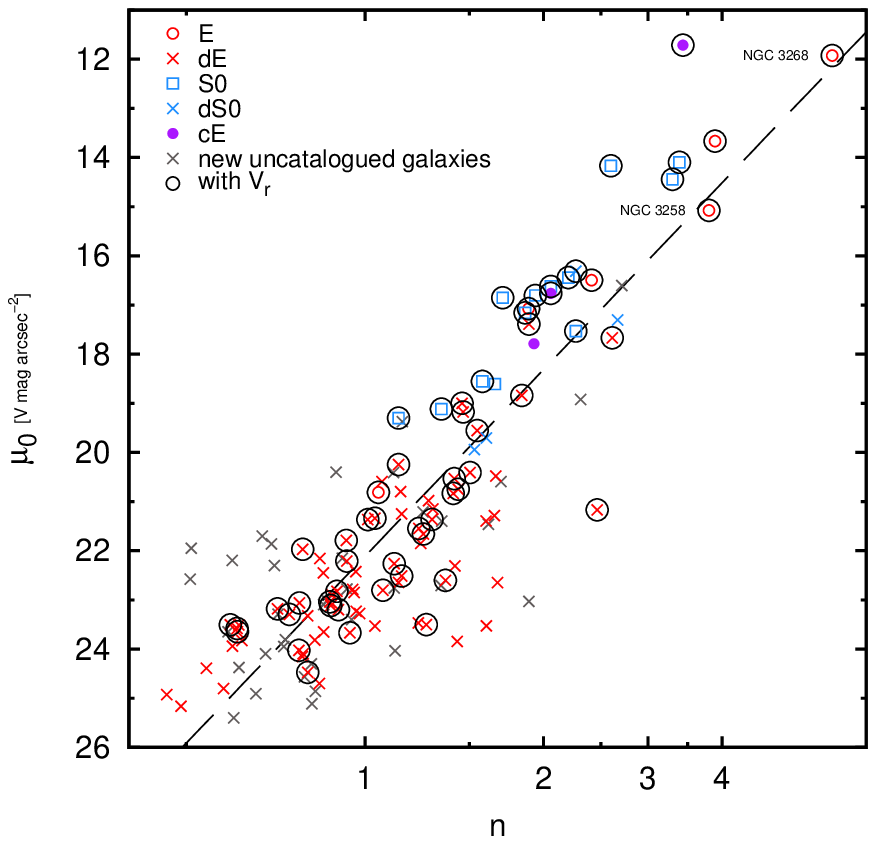}}\\
     \subfigure[Effective surface brightness versus absolute 
     magnitude ($V$-band).]{%
       \label{fig:Mv_mue}
       \includegraphics[width=\columnwidth]{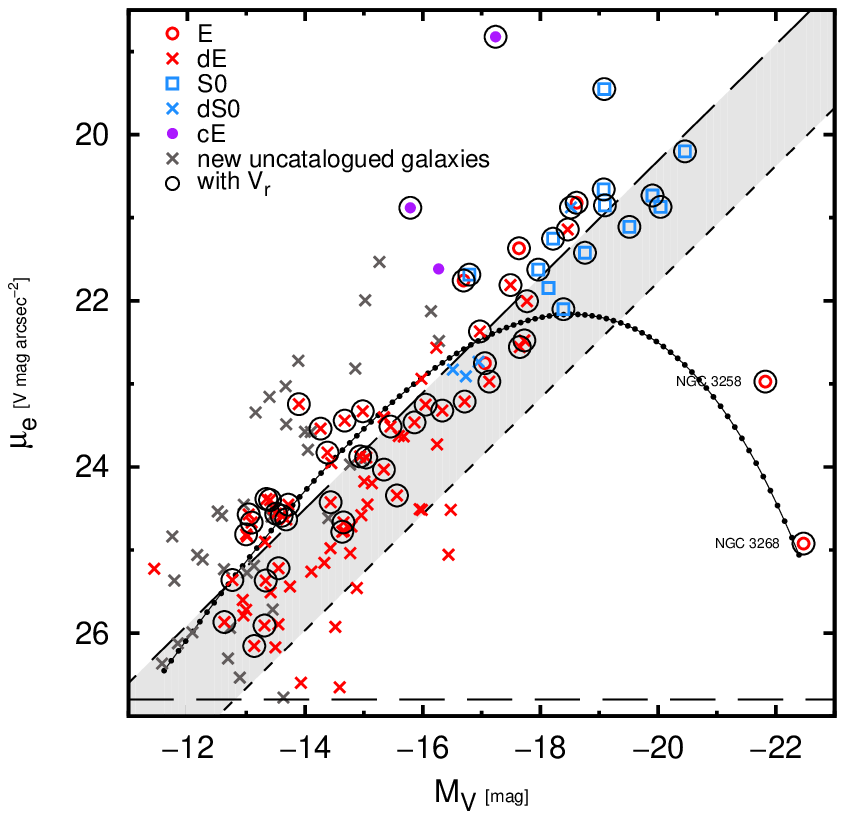}}%
     \subfigure[Logarithm of the effective radius versus absolute
     magnitude ($V$-band).]{%
       \label{fig:Mv_re}
       \includegraphics[width=\columnwidth]{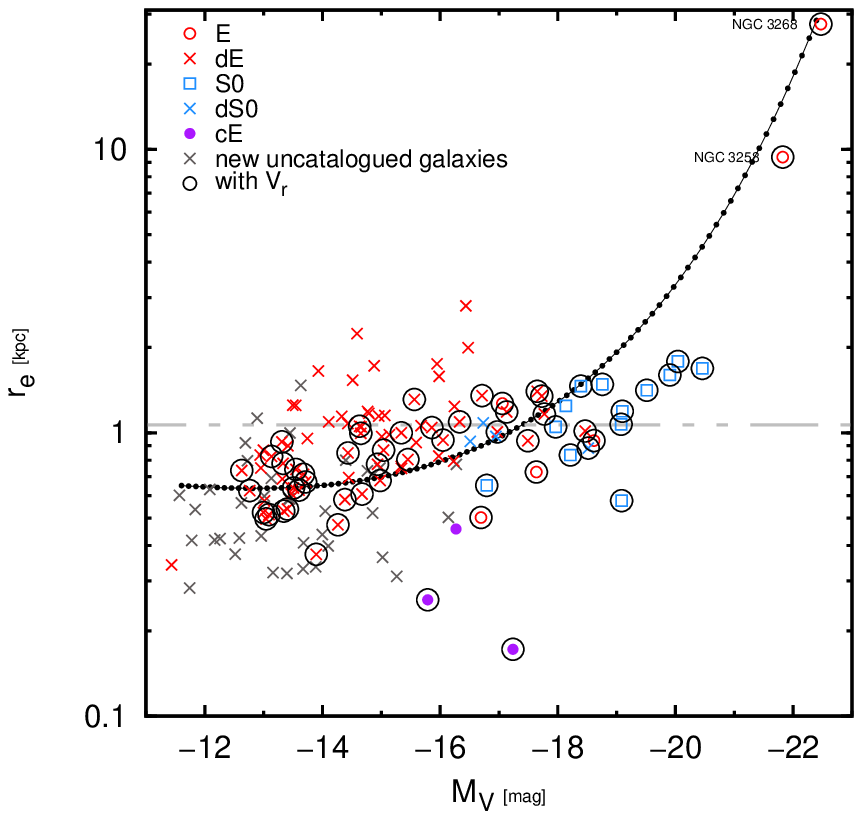}}%
     \caption{%
       Scaling relations for the Antlia final sample. Symbols are
       identified in each plot. Upper panels: dashed lines in plots
       (a) and (b) show the respective least--square linear fits,
         in plot (c) the dashed line corresponds to
         equation\,\protect\ref{eq:m0_n} (see text). Lower panels: dotted
       lines show the respective curved relations obtained following
       \protect\cite{2013pss6.book...91G}, the dashed lines in
       panel\,(d) refer to the completeness (see
       Sec.\,\ref{sec:completeness}), the horizontal dashed line in
       panel\,(e) indicates the dEs $\langle r_\mathrm{e} \rangle =
       1.07$\,kpc.}%
     \label{fig:scaling_relations}
  \end{center}
\end{figure*}

\subsection{Scaling relations involving structural
  parameters}\label{sec:scaling-relations-involving-structural-parameters}

Fig.\,\ref{fig:scaling_relations} shows five scaling relations for our
`final sample' of Antlia ETGs, namely panel\,(a) $\mu_0$ versus
absolute magnitude $M_V$, panel\,(b) $M_V$ versus
S\'ersic index $n$, and panel\,(c) $\mu_0$ versus $n$ at the top
row; panel\,(d) $\mu_\mathrm{e}$ versus $M_V$, and panel\,(e)
$r_\mathrm{e}$ versus $M_V$ at the bottom row. The different
symbols are explained within each plot, identifying morphological
types, new uncatalogued galaxies within $\pm 3 \sigma$ of the CMR,
and members confirmed with radial velocities. In order to provide an
easy comparison with results for other clusters, the parameters in
Washington $T_1$-band have been transformed into the $V$-band through
the equations given by \cite{1995PASP..107..945F} and the relation
$R_\mathrm{C} - T_1 \approx -0.02$ \citep{1996AJ....111..480G}.

The procedure outlined by \protect\cite{2013pss6.book...91G} was
followed with the aim of obtaining the expressions for the different
relations. The plots presented in Figs.\,\ref{fig:Mv_mu0} and
\ref{fig:Mv_n} roughly show linear relations followed by all the ETGs
except the two confirmed cEs. They extend from the gEs at the bright
end to the dSph candidates at the faint end, covering more than
10\,mag in $M_V$. The linear fits presented in these figures
were performed taking into account numerical errors in both axes and
excluding cE galaxies.  The following expressions are the respective
least-square linear fits, which are shown with dashed lines.
\begin{equation}\label{eq:Mv_mu0}
\mu_0 = (1.2 \pm 0.08)~M_V + (40.1 \pm 1.2)
\end{equation}
\begin{equation}\label{eq:Mv_n}
M_V =  (- 10.1 \pm 1.3)~\log(n) - (14.5 \pm 0.2)\textnormal{.}
\end{equation}

Equivalent equations are presented by \citet[and references
therein]{2013pss6.book...91G} for the data set compiled by
\cite{2003AJ....125.2936G}, in the $B$-band: $\mu_0 = 1.49~M_B +
44.03$ and $M_B = - 9.4~\log(n) - 14.3$ (we adopted $B-V = 0.96$ for E galaxies,
\citealt{1995PASP..107..945F}).

A linear relation between $\mu_0$ and $\log(n)$ is easily obtained
combining the two previous expressions,
\begin{equation}\label{eq:m0_n}
\mu_0 = -12.12~\log(n) + 22.70,  
\end{equation}
which is presented in Fig.\,\ref{fig:mu0_n} and provides a good match
to the whole set of Antlia data. This is expected as $\mu_0$ and $n$
are coupled variables in the S\'ersic model. The linear correlation
coefficients calculated for the relations depicted in panels
\ref{fig:Mv_mu0} and \ref{fig:Mv_n} give 0.9 and $-0.7$, respectively,
which indicates that the linear correlations provide good fits for all
of them.  Linear correlations $\mu_\mathrm{e} - M_V$ and $r_\mathrm{e}
- M_V$ are evident in the plots presented in Figs.\,\ref{fig:Mv_mue}
and \ref{fig:Mv_re}, followed by all the ETGs with the exception of
four confirmed members: the two dominant gEs (NGC\,3258 and NGC\,3268)
and the two cE or M32-type galaxies \citep[FS90\,110 and
FS90\,192,][]{2012MNRAS.419.2472S}. Lastly, a different way of fitting
these two latter relations is through the curved functions shown in
Figs.\,\ref{fig:Mv_mue} and \ref{fig:Mv_re} with dotted lines. It
should be noted that they follow from the empirical linear
correlations obtained before. Additionally, we have tested that small
changes in those linear equations lead to large variations in the
curved function in Fig.\,\ref{fig:Mv_mue} and small ones in
Fig.\,\ref{fig:Mv_re}.

In the next Section we will include a discussion relating the
different fits that can be performed to the scaling relations and how
they can be understood.

\subsection{The Kormendy relation}\label{sec:the-kormendy-relation}
The Kormendy relation (KR) is the correlation between mean effective
surface brightness and radius \citep{1985ApJ...295...73K}, more
precisely $\langle\mu_\mathrm{e}\rangle$ versus $\log(r_\mathrm{e})$, which
has proven to be a useful tool to study the formation of ETGs. 

\begin{figure*}
 \includegraphics[width=2.\columnwidth]{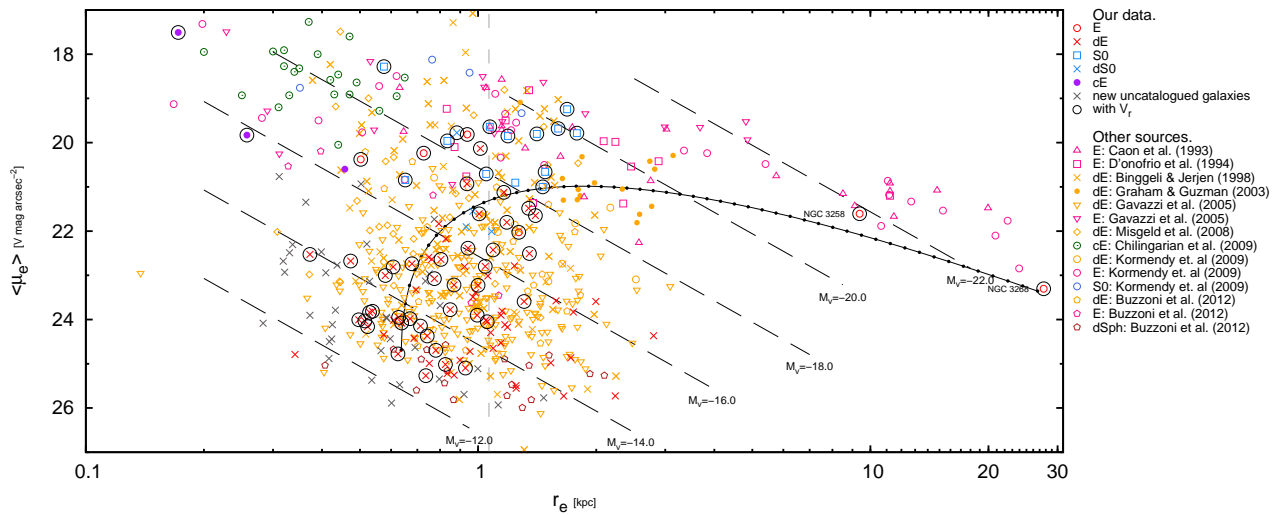}
  \caption{Kormendy relation for the Antlia final sample plus 
           data from other clusters and groups. Symbols are identified 
           on the right-hand side. The dotted line shows the curved 
           relation obtained with Antlia data following 
           \protect\cite{2013pss6.book...91G}, the vertical dashed 
           line indicates the $\langle r_\mathrm{e} \rangle$ for Antlia
           dEs, and the heavy long--dashed lines correspond to lines
           of constant $M_V$.}
  \label{fig:compara-re_muep}
\end{figure*}

Fig.\,\ref{fig:compara-re_muep} presents the KR for the Antlia
galaxies. Once more, bright and dwarf ETGs show different behaviours
on this plane. The bright ones follow a linear relation with the gEs
and cEs at opposite ends, while the dwarfs present a more disperse
distribution hard to disentangle.  On one hand, it has been proposed
that the distribution of dwarfs agree with the direction of the lines
of constant absolute magnitude, that are included in the plot. On the
other hand, we have seen that most dwarfs have a mean $r_\mathrm{e}$
close to 1\,kpc. This latter property is reflected by the curved
relation $\langle\mu_\mathrm{e}\rangle$ versus $r_\mathrm{e}$ obtained
by \cite{2011EAS....48..231G}.  In our case, this curved relation,
shown in Fig.\,\ref{fig:compara-re_muep}, was calculated using the
Antlia data by means of the linear relations depicted in
Figs. \ref{fig:Mv_mu0} and \ref{fig:mu0_n} excluding cE galaxies. Our
dwarf ETGs roughly follow this 1\,kpc mean value, although the
dispersion is quite high.

\subsection{Completeness}\label{sec:completeness}
It is important to clearly assess the completeness of the sample,
particularly when dealing with observations of faint objects. Since the main 
results of this paper are the numerical fits to scaling relations shown by a
sample of low surface brightness objects, completeness thereof directly
affects the fits and their dispersions.

The detection of the objects in the sample was performed visually,
which means that the images were inspected in detail, and all extended
objects displaying an elliptical morphology were identified. For this
reason, an automatic method to determine the completeness cannot be
applied. Following the procedure described by \citet[and references
therein]{2012A&A...538A..69L}, we used the effective surface
brightness-luminosity relation (Figure \ref{fig:Mv_mue}) for the final
sample to estimate its completeness limit. We performed a linear fit
of this relation and, in order to avoid any incompleteness effect, we
excluded from such fit all objects with $M_V > -16$\,mag, obtaining:
\begin{equation}\label{eq:Mv_mue_completeness}
\mu_{e} = (0.70 \pm 0.1)~M_{V} + (34.30 \pm 2.0) \textnormal{,}
\end{equation}
with a standard deviation $\sigma = 0.73$. On the other hand, taking
into account the sample's selection criteria (Section
\ref{sec:sample-selection}), we estimated the effective surface
brightness ($\mu_\mathrm{e,lim}$) for an object with S\'ersic index $n
= 1$ and effective radius $r_{e} = 1$\,kpc (typical for dEs), which
would show an isophotal radius $r = 10$\,arcsec at the $\mu_{(T_1)} =
27.5$\,mag\,arcsec$^{-2}$ isophote. The value obtained is
$\mu_\mathrm{e,lim\,(T_1)} = 26.2$\,mag\,arcsec$^{-2}$, which
corresponds to $\mu_\mathrm{e,lim} = 26.8$\,mag\,arcsec$^{-2}$ in the
$V$ band.

The long-dashed line in Figure\,\ref{fig:Mv_mue} shows the fit (from
Eq.~\ref{eq:Mv_mue_completeness}) and the short-dashed line shows
$\mu_{e} + 2 \sigma$ (shaded band), which defines a 96\% confidence
(or completeness). The line corresponding to the fainter $2 \sigma$
limit intersects with that representing $\mu_\mathrm{e,lim}=
26.8$\,mag\,arcsec$^{-2}$ at $M_{V} \approx -13.0$\,mag.  This means
that, supposing that the standard deviation remains constant along the
whole luminosity range, we have lost only $\sim 2\%$ of the faintest
galaxies at $M_{V} \approx -13.0$\,mag.

\section{Discussion}
The discussion on the photometric and structural scaling relations,  
that have been constructed for the final sample of the Antlia cluster, 
as well as a comparison with results for other clusters, will be given 
in this Section. 

\subsection{On the photometric relations}
The Antlia CMR of ETGs (Fig.\,\ref{fig:CT1_T1}) is a tight
correlation; along with similar relations in other clusters or groups,
it has been attributed mainly to a mass-metallicity relation \cite[and
references therein]{1999MNRAS.310..445T, 2012MNRAS.419.2472S}. CMRs
turned out to be an interesting property to study both from
observational and numerical simulations perspectives. From an
observational point of view, CMRs have been studied in many groups and
clusters \citep{1988ARA&A..26..509B, 1997PASP..109.1377S,
  2004MNRAS.349..527K}.  The `universality' of this relation for ETGs
has been suggested since the first studies, and many authors reported
a similar linear behaviour in different clusters
\citep{2007A&A...463..503M, 2008AJ....135..380L, 2008A&A...486..697M,
  2008MNRAS.386.2311S, 2009A&A...496..683M}, although the possible
existence of nonlinear trends has been analysed too
\citep{2006ApJS..164..334F, 2009ApJ...696L.102J}.

For instance, the existence of a break of the bright end of the
CMR, with respect to the linear fit performed on all the ETGs, has
already been discussed for about a decade
\citep[e.g.][]{2007AJ....133.1741B, 2011MNRAS.412..684B,
  2008ApJ...680..143G,
  2011EAS....48..231G}. \cite{2011MNRAS.417..785J} studied the bright
end of the CMR of galaxy clusters through a combination of
cosmological N-body simulations of clusters of galaxies and a
semi-analytic model of galaxy formation \citep{2008MNRAS.388..587L}.
In these simulations, this break of the bright end of the CMR
appears clearly as a consequence of galaxy evolution. The same effect
can be noticed in CMRs of observed ETGs, for instance at the Hydra\,I
cluster \citep{2008A&A...486..697M}, the Virgo cluster
\citep{2006ApJS..164..334F, 2009ApJ...696L.102J}, or a compilation
from the SDSS \citep{2009ApJ...699L...9S}. As noted by
\cite{2011EAS....48..231G}, it was already detectable in the study of
the Shapley\,8 galaxy cluster by \cite{1994MNRAS.267..431M}. In all of
them there is a `break' at the brighter magnitudes so that more
massive ETGs show almost constant colours, though different authors
give different interpretations. \cite{2011MNRAS.417..785J} explain
this behaviour as a consequence of the evolution of these bright
galaxies since redshift $z\sim 2$ being dominated by dry mergers, both
minor and major. In this case, galaxies would move towards brighter
magnitudes as they gain mass, while without gas no further star
formation (and enrichment) is expected, so their colours remain almost
invariable.

The absolute magnitude of this break for the Antlia cluster ($M_{T_1}
\approx -20$\,mag, see Fig.\,\ref{fig:CT1_T1}) is in agreement with
the value obtained by \cite{2011MNRAS.417..785J} for the simulated CMR
in the same photometric system, displayed in their fig.\,1. In the
same figure it can also be seen that the CMR is composed by galaxies
of increasing metallicity from the faint end to the bright end of the
relation. This latter result also supports the idea that metallicity
is the main responsible of the slope of the CMR.

Regarding the small scatter of the CMR \citep{1992MNRAS.254..601B,
  2001MNRAS.326.1547T, 2008MNRAS.386.2311S}, it is the consequence of
the scatter in both variables: age and metallicity, being the age of
the stellar population the prevailing one
\citep[e.g.][]{2006MNRAS.370.1106G}. On the one hand, this suggests
that the stellar population of ETGs has evolved passively since early
times. On the other hand, as another possible explanation, the scatter
measured with observations can be accounted for with models that
predict a continuous migration of late--type galaxies to the CMR, due
to different processes that stop their star formation
\citep{2009ApJ...695.1058R}.

\subsection{On the structural relations}
If we turn to the structural relations
(Fig.\,\ref{fig:scaling_relations}, panels\,(a) to \,(e)), the two gE
galaxies (NGC\,3258 and NGC\,3268) deserve a brief explanation in
order to understand their loci in such plots. They are both classified
as `core' profile E galaxies \citep[e.g. see][and references
therein]{2005A&A...440...73C, 2012AJ....143...78K}. Their profiles are
characterized by a shallow inner cusp, which is attributed to a
central deficit in luminosity with respect to the inward extrapolation
of the best-fitting global S\'ersic model. We cannot detect the cores
because they extend up to the inner 1\,arcsec and we fitted the
surface brightness profiles excluding the inner 1\,arcsec. Anyway, we
take into account the fact that the two gE galaxies are in fact `core'
ones. These `core' galaxies share other properties, like slow rotation
and boxy isophotes, in contrast to the fainter Es with `power-law'
profiles. Thus, if the central surface brightness $\mu_0$ of a `core'
E galaxy is taken from the cusp, it will have a lower (fainter) value
relative to the relation shown in panel\,(c) of
Fig.~\ref{fig:scaling_relations} and will fall well below it. However,
\cite{1997ASPC..116..239J} and \cite{2003AJ....125.2936G} noted that
if $\mu_0$ is taken from the inner extrapolation of the global
S\'ersic model, it will share the same linear relation with the other
ETGs. According to this, as the $\mu_0$ of the two Antlia gEs have
been calculated in this latter way, they should not be outliers of
these previous linear relations, within the scatter present in the
data.

It was indicated above that the data depicted in
panels\,\ref{fig:Mv_mue} and \ref{fig:Mv_re} follow linear
correlations, excepting four members: the two central gEs and the two
cEs. Regarding these four outliers, several authors have argued about
the existence of a dichotomy between `normal' and dwarf ETGs in
similar plots, meaning that data are placed along two different
sequences (or linear relations) separated by a gap at $M_V \approx
-18$ \citep[e.g.][and references therein]{ 2009ApJS..182..216K,
  2012ApJS..198....2K}.  According to this picture, gE and cE galaxies
are located at the opposite ends of a sequence defined by the `normal'
elliptical galaxies, being the cEs at the position of the brightest
$\mu_\mathrm{e}$ and smallest $r_\mathrm{e}$, while the other sequence
is mostly traced by the dEs. As reported in the Introduction, the
existence of these two `branches' is understood as an evidence that
they are distinct species.

In our plots at the bottom row of Fig.\,\ref{fig:scaling_relations}, it is 
difficult to establish the existence of separate sequences for non--dwarf
ETGs, mainly because in this magnitude range Antlia has S0 galaxies and very
few `normal' or bright Es. The gEs are clearly apart from the rest of ETGs,
but an alternative scenario is suggested by \citet[and references
therein]{2013pss6.book...91G}. Graham et al.  show that mathematical links
between the S\'ersic parameters as well as the empirical linear relations at
Figs.\,\ref{fig:Mv_mu0} and \ref{fig:Mv_n}, can be used to derive curved
relations for $\mu_\mathrm{e}$ versus $M_V$ and $\log(r_\mathrm{e})$ versus
$M_V$, that extend from dwarfs to gEs (e.g., fig. 12 in
\citealt{2003AJ....125.2936G}, and fig. 2.8 in
\citealt{2013pss6.book...91G}), without considering the cEs.  Such curved
relations are shown in Figs.\,\ref{fig:Mv_mue} and \ref{fig:Mv_re} with
dotted lines. In both plots, the loci of the two Antlia gEs are in very good
agreement with the respective curved relations, while at the opposite side
the curved relations match well with dEs. Both scaling relations seem to
connect dwarf and giant ellipticals and, according to this latter approach,
dwarfs appear to be the low-mass end of those sequences that unify the E
galaxies.  In this way, Es have a continuous range of concentrations,
measured by the S\'ersic shape index $n$ \citep{1993MNRAS.265.1013C}.  On
the other side, these curved functions do not seem to fit properly neither
the lenticular (particularly for panel\,(d)), nor the cE galaxies.  Galaxies
of these two types were not included in the analysis of scaling relations
performed by \cite{2013pss6.book...91G}.  However, Antlia S0s seem to fit
nicely when data from other systems are included (see below).

It must be taken into account that we are fitting the brightness
profiles of S0s and cEs with single S\'ersic models. Furthermore, due
to the few cEs and bright Es present in the Antlia cluster, we are unable
to test the \citet{2009ApJS..182..216K} scenario with our present
data.

It is interesting to note that the faintest Antlia galaxies, those
that look alike the dSph ones in the Local Group, seem to extend
almost all the scaling relations outlined in
Fig.\,\ref{fig:scaling_relations}, towards lower luminosities and
following the same trend as dEs. The only exception is the size, as
can be seen in Fig.\,\ref{fig:Mv_re}, where dSph candidates present
smaller effective radii than the mean $\langle r_\mathrm{e} \rangle =
1.07$\,kpc ($\sigma = 0.13$\,kpc). This average was calculated for the
dE galaxies ($-18 < M_V < -14$) in the final sample. That dSphs have
smaller radii than dEs has already been pointed out by
\cite{2012MNRAS.419.2472S} for fewer faint Antlia galaxies and
explained as a consequence of the limitations of the isophotal
photometry. In this paper, we calculate `total' integrated magnitudes
and surface brightnesses for all galaxies, although some
incompleteness is expected for galaxies fainter than $M_V \approx
-13$\,mag. Anyway, smaller effective radii are expected for faint
dwarfs, since any galaxy with an integrated magnitude $M_V \gtrsim
-13$ will necessarily have a small ($\lesssim 1$\,kpc) effecive
radius, unless its surface brightness be extremely (unphysically?)
low. On the other hand, selection effects would prevent against the
detection of any such galaxies; it can be shown that their isophotal
radii would fall below our limiting radius (see
Sect.~\ref{sec:completeness}).

A similar effect has been found for non-Antlia galaxies. For instance,
using a data compilation of a variety of stellar systems,
\cite{2013pss6.book...91G} presents a global analysis of sizes against
stellar masses.  It is visible in his fig.\,2.1 that dE galaxies have
half--light radii about 1\,kpc, while dSphs show a decline in the
sense that they have smaller radii as the stellar mass diminishes. A
quite similar figure has recently been presented by \citet[their
  figs.\,11 and 16]{2014MNRAS.443.1151N}. Another example of such a
different trend between the sizes of dEs and dSphs is given by
\cite{2008MNRAS.389.1924F}, in their fig\,7 that shows half--light
radius versus absolute $K$ magnitude for a different data set.

That dE galaxies seem to have an almost constant radius has already
been pointed out for several clusters and groups \citep[e.g.][and
references therein]{2008MNRAS.386.2311S}.  Our $\langle
r_\mathrm{e}\rangle$ is in agreement, among others, with that obtained
by \cite{2008A&A...486..697M} for the Hydra cluster, selecting
galaxies fainter than $M_V = -18$. This tendency is followed by the
curved relation shown in Fig.\,\ref{fig:Mv_re}, as the effective
radius tends to a constant value close to 1\,kpc at the faint end.

\cite{2009MNRAS.393..798D} show the scaling relations of an ETG sample
in different environments, including data of the Antlia cluster from
our first study \citep{2008MNRAS.386.2311S}. Their $\mu_\mathrm{0} -
M_{V}$ and $n - M_{V}$ diagrams extend along a larger range in
magnitude ($-8 \gtrsim M_{V} \gtrsim -24$\,mag).  The fits obtained in
this paper for both correlations (Figures \ref{fig:Mv_mu0} and
\ref{fig:Mv_n}, Equations \ref{eq:Mv_mu0} and \ref{eq:Mv_n})
reasonably agree within the magnitude range common to both samples.
\citeauthor{2009MNRAS.393..798D} remark that these scaling relations
show a change of slope at $M_{V} \sim -14$\,mag, that may be due to
different physical processes dominating the evolution of dEs and dSphs
but, due to our completeness limit, we cannot refer to such
low-luminosity range.

\subsection{Comparison with other galaxy clusters}\label{sec:comparison-with-other-galaxy-clusters}
\begin{figure*}
     \begin{center}
       \subfigure[Effective surface brightness versus absolute 
          magnitude ($V$-band).]{%
            \label{fig:comparision_Mv_mue}
            \includegraphics[width=2.\columnwidth]{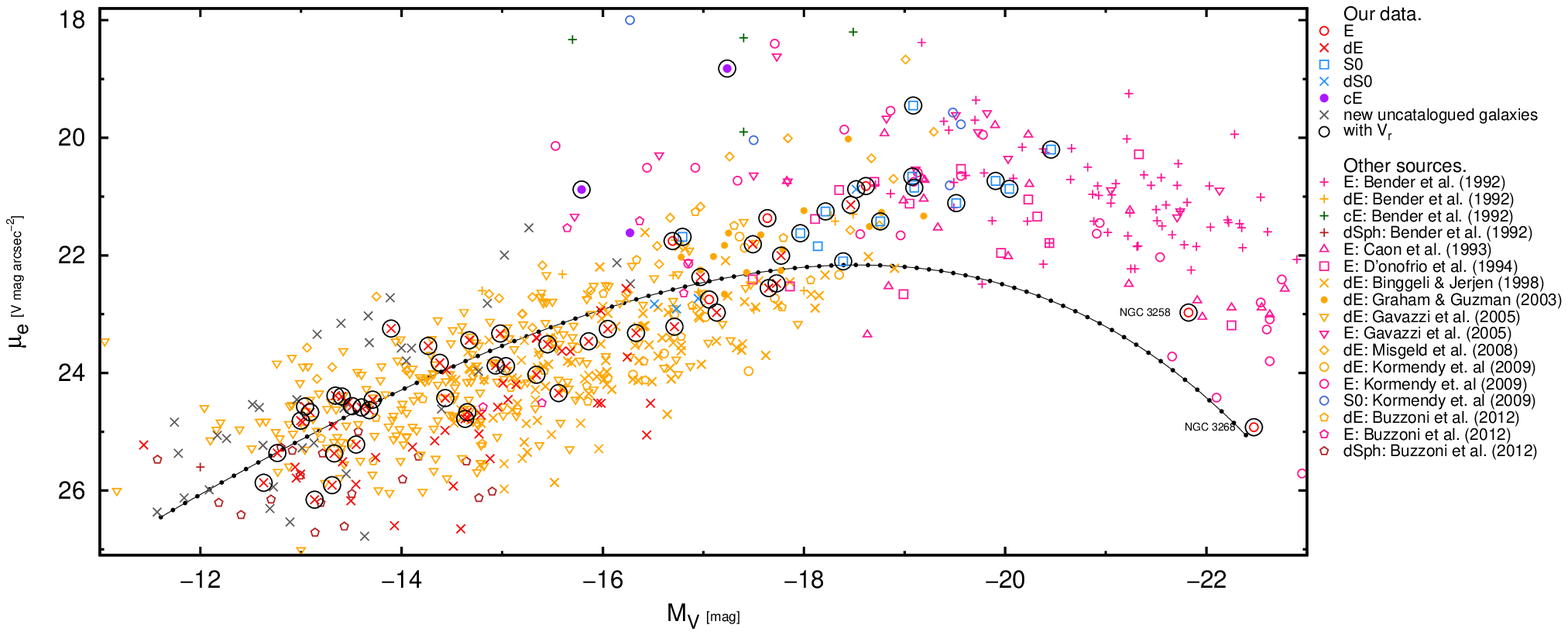}}\\
          \subfigure[Logarithm of the effective radius versus absolute 
             magnitude ($V$-band).]{%
            \label{fig:comparision_Mv_re}
            \includegraphics[width=2.\columnwidth]{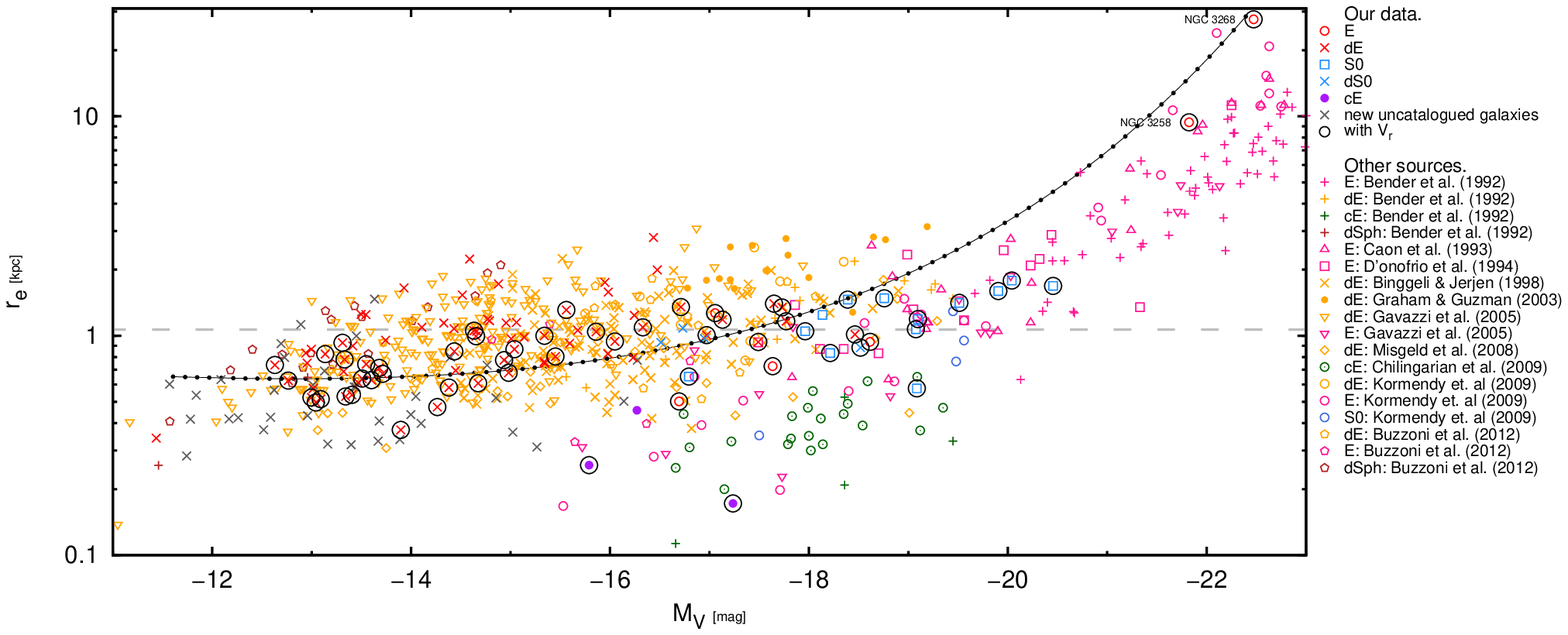}}
          \caption{%
            Scaling relations for the Antlia final sample plus data 
            from other clusters and groups. Symbols are identified on the 
            right-hand side of each panel. Lines have the same meaning 
            as in Fig.\,\ref{fig:scaling_relations}.}%
   \label{fig:comparision_scaling_relations}
    \end{center}
\end{figure*}

We took into consideration data from other galaxy clusters to better
understand the two scaling relations depicted at the bottom row of
Fig.\,\ref{fig:scaling_relations}, i.e. those that according to
\cite{2013pss6.book...91G} can be represented with curved functions,
and according to \citet{2009ApJS..182..216K} by two linear sequences
with different slopes.  In the seek of clarity, both graphs have been
reproduced and enlarged in
Fig.\,\ref{fig:comparision_scaling_relations}, adding the
corresponding parameters of ETGs from several groups and clusters,
like Fornax, Virgo, Coma, Hydra, and the NGC\,5044 group. Symbols and
the corresponding sources are identified in each panel. The curved
relations are the ones obtained in the previous section using the
Antlia data (Figs.~\ref{fig:Mv_mue} and \ref{fig:Mv_re}).

Fig.\,\ref{fig:comparision_Mv_mue} shows that the Antlia ETG data
follow the same trend as those from other systems, preserving a
similar dispersion and in good agreement with the fainter end of the
curved relation. The Antlia cEs share the same locus as other cEs and,
as expected, they all have higher $\mu_\mathrm{e}$ than ETGs of
similar luminosities. The two Antlia gEs are located close to the
bright end of the curved relation, as in fact it was calculated for
Antlia, but most bright Es ($M_V < -18$) from other systems have
brighter $\mu_\mathrm{e}$ and form a kind of parallel sequence above
the curved function. On the one hand, the ETGs seem to be distributed
in this plot along a single sequence that joins dwarfs, S0s, and
normal/bright ellipticals, being the cEs the only ones that depart
from it. On the other hand, the dispersion is quite different for ETGs
brighter and fainter than $M_V \sim -18$ and the existence of two
different linear sequences cannot be discarded, particularly owing to
the locus of the cE galaxies.

Due to the links between the different parameters derived from
S\'ersic models, a similar analysis applies to the size--luminosity
relation presented in Fig.\,\ref{fig:comparision_Mv_re}. Antlia ETGs
share the same locus as equivalent galaxies from other clusters.  The
Antlia curved relation is roughly applicable to part of the ETGs,
leaving out the cEs, which have smaller radii than ETGs of similar
luminosity.  The brightest Es at the large--size extreme and the dSphs
at the opposite end, also fall below the curved relation. Once more,
no `gaps' are visible in this size--luminosity relation, but the
existence of two linear sequences with different slopes cannot be
dismissed now. In comparison with the plot in panel\,(a), more cEs
have available data to be included in this graph
\citep{2009Sci...326.1379C} and it looks much more likely that giant
Es and cEs may fall on the same sequence, while the other sequence
involves mainly the dEs.

\cite{2008ApJ...689L..25J} performed a study of Virgo ETGs with
homogeneus data from the Sloan Digital Sky Survey (SDSS) and concluded
that giants and dwarfs do not form one single sequence in the
size--luminosity relation.  On the other side,
\citet{2006ApJS..164..334F} carried out an analysis of Virgo ETGs with
homogeneus data obtained with the Advanced Camera for Surveys of the
Hubble Space Telescope (ACS--HST) and gave support to the alternative
scheme, though they suggested that dEs may be a diverse population
with different origin and evolution.\nocite{1992ApJ...399..462B,
  1993MNRAS.265.1013C, 1994MNRAS.271..523D, 1998A&A...333...17B,
  2003AJ....125.2936G, 2005A&A...430..411G, 2008AJ....136..773F,
  2008A&A...486..697M, 2009ApJS..182..216K, 2012MNRAS.420.3427B,
  2012A&A...538A..69L}

With the aim of comparison and to fill empty regions in the KR
diagram, we have added in Fig.\,\ref{fig:compara-re_muep} data from
other clusters and groups, like in Fig.\,\ref{fig:comparision_Mv_re}.
The KR can be physically understood as a projection, on the surface
photometry plane, of the fundamental plane (FP) of ETGs
\citep{1987ApJ...313...59D}. The FP links
$\langle\mu_\mathrm{e}\rangle$, $\log(r_\mathrm{e})$, and
$\log(\sigma_0)$, being $\sigma_0$ the central velocity dispersion. It
was originally derived for bright early-type galaxies, i.e. those with
$M_B \lesssim -18$, but there seems to be no unique KR for bright and
dwarf ETGs. For instance, \cite{2008ApJ...685..875D} studied the FP of
a sample of more than 1500 ETGs and concluded that the coefficients of
the FP, and accordingly those of the KR, depend on the absolute
magnitude range of the galaxy sample. On the basis of ETGs in 16
nearby groups, \cite{2004MNRAS.349..527K} show that the KRs followed
by bright and dwarf ETGs are offset and have different slopes. It can
be seen in their fig. 6 that the KR outlined by the dwarfs is aligned
with the lines of constant absolute
magnitude. \cite{2005A&A...438..491D} also show, through a kinematical
study of a sample of dEs in groups and clusters, some structural
differences between bright and dwarf ETGs considering different
projections of the FP.  For example, in the log Re - log Ie relation
(their fig. 1d) different slopes can be seen for bright ellipticals
and for dEs. The dE galaxies have Re close to 1 kpc, with a large
scatter, while bright and intermediate luminosity Es show a linear
relation with quite a different slope, in a similar way as the KR we
have obtained. The authors note that the correlations the dEs follow
(as the KR) are not as tight as in the case of the bright ellipticals,
probably due to the higher sensitivity of the low-mass galaxies to
internal and external procceses (supernova explosions, feedback
efficiency, galactic winds, tidal stripping, ram-pressure stripping of
gas, etc). In particular, the low-mass galaxies (dEs and dShps) lie
above the FP defined by the bright elliptical galaxies of their
sample.

\cite{2013pss6.book...91G} explains the original KR for just bright ETGs
as: `a tangent to the bright arm of what is actually a curved
distribution'. He also sustains that the different trend of the
dwarfs in this relation, that maintain an almost constant radius, does
not imply that different physical mechanisms are at work in bright and
dwarf ETGs, but that they follow a continuous structural variation
that depends on the shape of the brightness profiles with luminosity.

We are again faced to this dualism. The results deduced from the
Antlia data seem to be more coincident with the \citet[and references
  therein]{2013pss6.book...91G} proposal, i.e.  the existence of
unique relations with a continuous variation from bright to dwarf
ETGs, but excluding the cEs. When we add data from other systems the
situation is not clear.  If the cEs are considered jointly with
the bright ETGs, the evidence of the existence of two distinct families 
like in the scenario supported by \citet[and references 
therein]{2009ApJS..182..216K} seems more appropriate.

\section{Summary and Conclusions}\label{sec:summary-and-conclusions}
We have presented the first large--scale analysis of the photometric and
structural scaling relations followed by the early-type galaxy population of
the Antlia cluster.  These relations were built on the basis of surface
photometry performed on MOSAIC\,II--CTIO images for 177 ETGs, being 44 per
cent of them newly discovered ones. Out of this ETG sample, 53 galaxies are
members confirmed through radial velocities, measured on new GMOS--GEMINI
and VIMOS--VLT spectra, as well as obtained from the literature. The ETGs
that lack spectra have high probability of being members due to the
membership status 1 assigned by FS90 (`definite' members) and/or because
their photometry places them within $\pm 3 \sigma$ of the CMR
(Fig.\,\ref{fig:CT1_T1}).

Total integrated magnitudes and colours in addition to accurate
structural parameters were obtained, for every galaxy, by fitting
single S\'ersic models to the observed surface brightness profiles and
integrating them to infinity. Based on them, we constructed the
scaling relations for the Antlia cluster.

The colour--magnitude plane in the Washington photometric system shows
that all ETGs follow a tight linear relation, spanning more than
10\,mag. Almost all galaxies with spectroscopically confirmed
membership lie within $\pm 3 \sigma$ of the CMR, giving thus support
to the cluster as an entity.  A break at the bright end of the
CMR is discernible, which is understood as a consequence that dry
mergers dominate the formation of gE galaxies since $z \sim 2$
\citep[e.g.][]{2011MNRAS.417..785J}.

Linear relations can be fitted to the Antlia ETGs in the planes $\mu_0
- M_V$ and $M_V - \log(n)$. Following the procedure explained by
\citet[and references therein]{2013pss6.book...91G}, these linear
relations plus the equations that link the S\'ersic parameters, let us
derive two curved functions that match most ETGs in the planes
$\mu_\mathrm{e}- M_V$ and $\log(r_\mathrm{e}) - M_V$. The two Antlia
confirmed cE galaxies do not follow those curved relations. Most of
the S0 galaxies are also outliers of the curved relation
$\mu_\mathrm{e} - M_V$, too. Due to the few bright Es and cEs present
in Antlia, it is not possible to compare whether two linear relations,
with different slopes, can be fitted in these latter planes instead of
the curved functions \citep[and references
therein]{2009ApJS..182..216K}. We remind that brightness profiles for
the cE and S0 galaxies have been fitted with single S\'ersic models.

When data from other clusters and groups are included in the planes
$\mu_\mathrm{e} - M_V$ and $\log(r_\mathrm{e}) - M_V$, bright Es and
cEs fill in the almost empty regions. In these cases, like for the
Kormendy relation, a match with two different linear relations for
bright and dwarf ETGs is a valid option.  The curved relations derived
previously with just the Antlia data, provide a reasonable match if
cEs are left aside, though the match for the brighter galaxies is
offset.

We plan to continue our study of Antlia, extending our coverage to encompass
the whole cluster, in order to build the scaling relations including the
entire galaxy population. Clearly, a set of homogeneus data from which a
careful derivation and fit of the observed brightness profile of every
galaxy are obtained, is the unavoidable first step to settle which scenario
is more appropriate to better explain the scaling relations. From this
starting point, a deeper structural analysis of the different galaxy types,
along with their stellar populations and spatial distribution is needed.

\section*{Acknowledgements}
We thank the referee Dr. Thorsten Lisker for valuable comments that
helped to improve this manuscript. This work was funded with grants
from Consejo Nacional de Investigaciones Cient\'ificas y T\'ecnicas de
la Rep\'ublica Argentina, Agencia Nacional de Promoci\'on Cient\'ifica
y Tecnol\'ogica, and Universidad Nacional de La Plata
(Argentina). JP~Calder\'on, LPB and JP~Caso are grateful to the
Departamento de Astronom\'ia de la Universidad de Concepci\'on (Chile)
for financial support and warm hospitality during part of this
research. TR acknowledges financial support from FONDECYT project
Nr. 1100620, and from the BASAL Centro de Astrofisica y Tecnologias
Afines (CATA) PFB-06/2007. LPB and MG: Visiting astronomers,
  Cerro Tololo Inter-American Observatory, National Optical Astronomy
  Observatories, which are operated by the Association of Universities
  for Research in Astronomy, under contract with the National Science
  Foundation. The present paper is also based on observations made
  with ESO telescopes at the La Silla Paranal Observatory under
  programme ID 79.B-0480, and obtained at the Gemini Observatory,
  which is operated by the Association of Universities for Research in
  Astronomy, Inc., under a cooperative agreement with the NSF on
  behalf of the Gemini partnership: the National Science Foundation
  (United States), the National Research Council (Canada), CONICYT
  (Chile), the Australian Research Council (Australia), Minist\'erio
  da Ci\^encia, Tecnologia e Inova\c{c}\~ao (Brazil), and Ministerio
  de Ciencia, Tecnolog\'ia e Innovaci\'on Productiva (Argentina).
\bibliographystyle{mn2e}

\bibliography{Calderon2015}

\bsp
\label{lastpage}

\end{document}